\documentclass[12pt]{article}
\usepackage{amsmath}
\usepackage{graphicx,psfrag,epsf}
\usepackage{enumerate}
\usepackage[authoryear, round]{natbib}
\usepackage{comment}
\usepackage[toc,page]{appendix}
\usepackage[color]{changebar}
\cbcolor{white}

\newcommand{\blind}{0}

\usepackage{amsfonts}
\usepackage[colorlinks=true,linkcolor=blue, citecolor=blue]{hyperref}%
\usepackage{xcolor}
\definecolor{applegreen}{rgb}{0.55, 0.71, 0.0}

\begin{document}

\def\spacingset#1{\renewcommand{\baselinestretch}%
{#1}\small\normalsize} \spacingset{1}


\if0\blind
{
    \date{}
  \title{\bf Bayesian Hyperbolic Multidimensional Scaling}
  \author{Bolun Liu\thanks{
    The authors gratefully acknowledge support from NIH grants R01 HD070936 (Raftery) and DP2 MH122405 (McCormick).  Correspondence: tylermc@uw.edu.}\hspace{.2cm}\\
    Department of Biostatistics\\ Bloomberg School of Public Health\\
    Johns Hopkins University
    \and ~\\
    Shane Lubold \\
    Department of Statistics\\ University of Washington
    \and ~\\
    Adrian E. Raftery \\
    Department of Statistics\\
    Department of Sociology\\ University of Washington
    \and ~\\
    Tyler H. McCormick \\
    Department of Statistics\\
    Department of Sociology\\ University of Washington}
  \maketitle
} \fi

\if1\blind
{
  \bigskip
  \bigskip
  \bigskip
  \begin{center}
    {\LARGE\bf Title}
\end{center}
  \medskip
} \fi

\bigskip
\begin{abstract}
Multidimensional scaling (MDS) is a widely used approach to representing high-dimensional, dependent data. MDS works by assigning each observation a location on a low-dimensional geometric manifold, with distance on the manifold representing similarity.  We propose a Bayesian approach to multidimensional scaling when the low-dimensional manifold is hyperbolic.  Using hyperbolic space facilitates representing tree-like structures common in many settings (e.g. text or genetic data with hierarchical structure).  A Bayesian approach provides regularization that minimizes the impact of  measurement error in the observed data and assesses uncertainty.  We also propose a case-control likelihood approximation that allows for efficient sampling from the posterior distribution in larger data settings, reducing computational complexity from approximately $O(n^2)$ to $O(n)$.  We evaluate the proposed method against state-of-the-art alternatives using simulations, canonical reference datasets, Indian village network data, and human gene expression data. Code to reproduce the result in the paper is available at \url{https://github.com/peterliu599/BHMDS}.
\end{abstract}

\noindent%


\spacingset{1.45}
\section{Introduction}
\label{sec:intro}

Multidimensional scaling (MDS) methods represent high-dimensional data in a low-dimensional space, using dissimilarities between the observations as a means of identifying positions \citep{Kruskal}.  Observations with small dissimilarities will be placed close together, while those with larger dissimilarities will be placed further apart. A long literature on MDS methodology illustrates the utility of MDS as a means of summarizing complex, dependent data and for downstream applications, such as detecting clusters from the dissimilarities \citep{Borg, davison, cox}. 

In many settings, the observed dissimilarities we wish to apply MDS methods to are likely to contain measurement error.  These errors could arise from misreporting in the context of the social sciences (e.g. a retrospective behavioral inventory) or from miscalibration of machinery or operator error in industrial settings.  Using a probabilistic model is one way to account for this additional uncertainty.   \cite{Takane, Groenen_93, Mackay} and others have proposed maximum likelihood MDS methods for handling measurement error. The use of these methods relies on asymptotic theory, which might not apply to sample sizes used in applications, and the problem requires solving a non-linear optimization problem where the number of parameters grows quickly as the sample size grows \citep{cox}.  One potential framework to address these potential issues with MDS is a Bayesian framework. \cite{oh2001bayesian} provided a Bayesian procedure to estimate the configuration of objects given (potentially noisy) dissimilarities between objects. Extensions of Bayesian MDS to the case of large datasets were discussed in \cite{Holbrook}.

Along with the statistical framework, another critical, but often ignored, choice in implementing MDS is the choice of geometry for the low-dimensional manifold.  Multidimensional scaling methods often assume the observed dissimilarities are computed using Euclidean distances among objects in a Euclidean space. \cite{oh2001bayesian}, for example, assume a Euclidean distance model with a Gaussian measurement error and propose a Markov-Chain Monte Carlo algorithm to compute a Bayesian solution for the object configuration.  Yet there is a growing literature showing that representing objects in other embedding spaces might lead to better representations and therefore be more useful in downstream tasks \citep{tradeoff, Re_Knowledge, smith2019geometry, Lubold2020, hydra}. In particular, hyperbolic spaces, defined in Section \ref{sec: meth}, have been shown to produce embeddings with lower distortion, especially for data that is hierarchical or tree-like.

\cbstart
In this work, we combine the hyperbolic MDS methods with a Bayesian procedure. Specifically, we apply the Bayesian MDS procedure from \cite{oh2001bayesian} to a hyperbolic space. Our model posits that the objects of interest are located in positions on a low-dimensional hyperbolic space, with the true dissimilarities between objects being the hyperbolic distances between positions. We further assume that the observed dissimilarities are the true dissimilarities plus a Gaussian error.
 We then derive a Markov-chain Monte Carlo (MCMC) method we use to obtain Bayesian estimates of the object positions, which allows practitioners to calculate distance-based measures (e.g. distances between positions, distances to the hyperbolic origin, etc.) that are often the pragmatic goals in applications.  
\cbend

The paper is organized as follows. In Section \ref{sec: meth}, we formally define the Hyperbolic space model used in this work and posit a model for observed dissimilarities computed from points in Hyperbolic space. We then discuss a prior distribution over this space and derive an MCMC algorithm to draw samples from the posterior in Section \ref{sec: posterior}. Sections \ref{sec: simulations} and \ref{sec: data_analysis} contain simulations and applications of our method to real datasets in genomics. We conclude in Section \ref{sec: conclusion}.

\section{Hyperbolic Geometry}

We now discuss the mathematical details of the hyperbolic geometry. The hyperbolic geometry is a non-Euclidean geometry that has a constant negative curvature and is commonly visualized as the upper sheet of the unit two-sheet hyperboloid.
There exist multiple equivalent hyperbolic models, such as the Klein model, the Poincaré disk model, and the Lorentz (Hyperboloid/Minkowski) model. We use the Lorentz model to parameterize the hyperbolic geometry, which parallels the representation used in existing hyperbolic MDS algorithms \citep{hydra, tradeoff}.  This representation also facilitates convenient priors for our Bayesian model in Section \ref{sec: meth}.

To define the Lorentz model,  we begin with the definition of the Lorentzian product. For any $\boldsymbol x = (x_0, \dotsc, x_p)$ and $\boldsymbol  y = (y_0, \dotsc, y_p) \in \mathbb{R}^{p + 1}$, the Lorentzian product $  \left\langle\boldsymbol{x}, \boldsymbol{y}\right\rangle_{\mathcal{L}}$ is defined as 
\begin{equation*}
    \left\langle\boldsymbol{x}, \boldsymbol{y}\right\rangle_{\mathcal{L}} \equiv -x_{0} y_{0} +\sum_{i=1}^{p} x_{i} y_{i} \;.
\end{equation*}
The $p$-dimensional Lorentz model with curvature $-\kappa$, which we denote by $\mathbb{H}^p(\kappa)$, can be represented as a collection of coordinates $\boldsymbol x \in \mathbb{R}^{p+1}$ with $x_0 > 0$ such that its Lorentzian product with itself is $-1$ and equipped with the hyperbolic distance proportional to the square root of $\kappa$. That is,
\begin{equation*}
    \mathbb{H}^p(\kappa) \equiv \left\{\boldsymbol x \in \mathbb{R}^{p + 1}: x_0 > 0, \langle\boldsymbol{x}, \boldsymbol{x}\rangle_{\mathcal{L}}= -1 \right\}, \quad \kappa > 0\;,
    \label{eq: def_alter}
\end{equation*}
equipped with the hyperbolic distance 
\begin{equation}
    d_{\mathbb{H}^{p}(\kappa)}(\boldsymbol x, \boldsymbol y) \equiv \frac{1}{\sqrt{\kappa}} \operatorname{arccosh} \left( -\left\langle\boldsymbol{x}, \boldsymbol{y}\right\rangle_{\mathcal{L}}\right) \;,
    \label{eq: dist}
\end{equation}
which is the geodesic distance between $\boldsymbol x$ and $\boldsymbol y$ on $\mathbb{H}^p(\kappa)$. Specifically, the curvature $-\kappa$ controls the hyperbolicity of the geometry so that the space becomes more hyperbolic as $-\kappa$ becomes more negative and becomes flatter as $\kappa$ shrinks to zero (Euclidean geometry has curvature exactly 0).

\section{Bayesian Modeling Framework}
\label{sec: meth}

We now describe our statistical framework for Bayesian Hyperbolic Multidimensional Scaling (BHMDS), which represents the objects of interest by coordinates in hyperbolic geometry, so that the hyperbolic distances between objects resemble their true dissimilarity measures. We suppose the objects dwell on $\mathbb{H}^p(\kappa)$ with coordinates $\boldsymbol x_1, \boldsymbol x_2, \cdots, \boldsymbol x_n$, and denote by $\delta_{ij}$ the dissimilarity measure between object $i$ and object $j$ as well as the hyperbolic distance between $\boldsymbol x_i$ and $\boldsymbol x_j$:
\begin{equation}
    \delta_{ij} \equiv d_{\mathbb{H}^{p}(\kappa)}(\boldsymbol x_i, \boldsymbol x_j) = \frac{1}{\sqrt{\kappa}} \arccos \left( -\left\langle\boldsymbol{x_i}, \boldsymbol x_j\right\rangle_{\mathcal{L}}\right),\quad i, j = 1, \cdots, n \;.
    \label{true_mea}
\end{equation}

In many settings, the observed dissimilarities contain measurement errors. As in \cite{oh2001bayesian}, we represent the observed dissimilarity $d_{ij}$ as the true dissimilarity plus a Gaussian error, with the constraint that the observed dissimilarity is always positive. We therefore assume that $d_{ij}$ follows a truncated normal distribution:
\begin{equation}
    d_{i j} \sim N\left(\delta_{i j}, \sigma^{2}\right) I\left(d_{i j}>0\right), \quad i < j, \quad i, j=1, \ldots, n \; ,
    \label{eq:model}
\end{equation}
where $\delta_{ij}$ is as defined in (\ref{true_mea}), $\boldsymbol x_i$ are unobserved, and $\sigma^2$ is the variance of the measurement error. 

Given the statistical model, we now specify priors for $\boldsymbol{X} = \{\boldsymbol x_1, \cdots, \boldsymbol x_n\}$ and $\sigma^2$. First note that, using (\ref{eq:model}), the likelihood of the unknown parameters $\boldsymbol X$ and $\sigma^2$ is
\cbstart
\begin{equation}
    l\left(D \; | \; \mathbf{X}, \sigma^{2} \right) \propto\left(\sigma^{2}\right)^{-m / 2} \exp \left[-\frac{1}{2 \sigma^{2}} S S R-\sum_{i<j} \log \Phi\left(\frac{\delta_{i j}}{\sigma}\right)\right] \;,
    \label{eq:lik}
\end{equation}
\cbend
where $m=n(n-1) / 2$ is the number of dissimilarities, $S S R=\sum_{i<j}\left(\delta_{i j}-d_{i j}\right)^{2}$ is the sum of squared residuals, $\Phi$ is the cumulative distribution function of the standard normal random variable, and $D = \{d_{ij}\}_{i, j = 1, \cdots, n}$ is the matrix of observed dissimilarities.  The square root of the $S S R$ term in the likelihood is often referred to as \emph{stress} in the MDS literature, meaning that our approach falls under the umbrella of stress-minimizing approaches to MDS.

\cbstart
We now elaborate on the prior choice for hyperbolic coordinates. Due to the constraint in (\ref{eq: def_alter}), hyperbolic coordinate distributions are generally asymmetric and complex in form, often leading to intractable full conditional posterior distributions. In such settings, Metropolis-Hasting samplers are computationally challenging. Therefore, the choice of a prior structure has substantial implications for the feasibility of posterior sampling. 
We use the hyperbolic wrapped normal distribution centered on the hyperbolic origin~\citep{wrapnorm} as our hyperbolic prior. This prior distribution equalizes sampling on $\mathbb{H}^p$ to sampling on $\mathbb{R}^p$ by defining a one-to-one transformation from $\mathbb{R}^p$ to $\mathbb{H}^p$. Specifically, to sample from the hyperbolic prior, we first sample $\boldsymbol{v} \sim \mathcal{N}_p(\boldsymbol{0}, \Lambda) \in \mathbb{R}^p$, then transform $\boldsymbol v \in \mathbb{R}^p$ to $\boldsymbol x \in \mathbb{H}^p$ by transformation $T(\cdot)$ specified by the prior,
\begin{equation}
    \boldsymbol x = T(\boldsymbol v) = \cosh \left(\|\Tilde{ \boldsymbol{v}}\|_{\mathcal{L}}\right) \boldsymbol{\mu}_0^p+\sinh \left(\|\Tilde{\boldsymbol{v}}\|_{\mathcal{L}}\right) \frac{\Tilde{\boldsymbol{v}}}{\|\Tilde{\boldsymbol{v}}\|_{\mathcal{L}}} \;,
    \label{eq:map}
\end{equation}
where $\boldsymbol \mu_0^p= (1, 0, \cdots, 0) \in \mathbb{H}^{p}$ is the hyperbolic origin, $\Tilde{\boldsymbol v} = (0, \boldsymbol v) \in \mathbb{R}^{p+1}$, and $\|\Tilde{ \boldsymbol{v}}\|_{\mathcal{L}} \equiv \sqrt{\langle\Tilde{ \boldsymbol{v}}, \Tilde{ \boldsymbol{v}}\rangle_{\mathcal{L}}}$. The procedure produces, $\boldsymbol{x}$, which is a sample from the hyperbolic wrapped normal distribution. The transformation $T(\cdot)$ enables us to reparameterize the likelihood in (\ref{eq:lik}) using $\boldsymbol{V} = \{\boldsymbol{v}_1, \cdots, \boldsymbol{v}_n\}$, such that 
\begin{equation}
    l\left(D \; | \; \mathbf{V}, \sigma^{2} \right) = l\left(D \; | \; \mathbf{X}, \sigma^{2}  \right) \propto\left(\sigma^{2}\right)^{-m / 2} \exp \left[-\frac{1}{2 \sigma^{2}} S S R-\sum_{i<j} \log \Phi\left(\frac{\delta_{i j}}{\sigma}\right)\right] \;,
    \label{eq:lik_new}
\end{equation}
where $\delta_{ij} = d_{\mathbb{H}^{p}(\kappa)}\left(T(\boldsymbol v_i) , T(\boldsymbol v_j)\right)$. Imposing a Gaussian prior on $\boldsymbol{V}$ is equivalent to a hyperbolic wrapped normal prior on $\boldsymbol{X}$. The transformation allows us to sample in the space of the Gaussian prior, meaning we can use a simple random walk Metropolis-Hasting algorithm to sample from the posterior.  
\cbend

We next define the prior for the observation error variance, $\sigma^2$.  We use a conjugate prior $\sigma^2 \sim IG(a, b)$, the Inverse Gamma distribution with mode $b / (a + 1)$. For the hyperprior on the diagonal entries of the auxiliary variable variance matrix $\Lambda=\operatorname{Diag}\left(\lambda_{1}, \ldots, \lambda_{p}\right)$, we also assume a conjugate Inverse Gamma prior, $\lambda_{j} \sim I G\left(\alpha, \beta_{j}\right)$, independently for $j = 1, \cdots, p$. We will further assume prior independence among $\boldsymbol V$, $\Lambda$, and $\sigma^2$, i.e., $\pi\left(\mathbf{V}, \sigma^{2}, \Lambda\right)=\pi(\mathbf{V}) \pi\left(\sigma^{2}\right) \pi(\Lambda)$, where $\pi(\mathbf{V}), \pi\left(\sigma^{2}\right)$, and $\pi(\Lambda)$ are the priors given earlier.

\cbstart
Often there is little prior information about $\boldsymbol V$, $\Lambda$, and $\sigma^2$. \cite{oh2001bayesian} proposed using vague prior distributions centered around frequentist MDS estimates. We follow a similar strategy here. Specifically, we use the embedding result from \texttt{hydraPlus}, the stress-minimizing hyperbolic MDS algorithm proposed in \cite{hydra}, to choose the parameters of the prior distributions. For the hyperparameters of $\sigma^2$, one can set a small value of $a$, e.g. $a = 5$, for a vague prior of $\sigma^2$, and choose $b = (a - 1) SSR^{(0)} / m$, where $m = n(n-1) / 2$, and $SSR^{(0)}  =  \sum_{i<j}\left(\delta_{i j}^{(0)}-d_{i j}\right)^{2}$
is the sum of squared residuals calculated using \texttt{hydraPlus} embedding result, so that the prior mean matches $SSR^{(0)} / m$. Similarly, for the hyperprior distribution of $\lambda_j$, we choose a small value of $\alpha$, e.g. $\alpha = 0.5$, and choose $\beta_j$ such that the prior mean of $\lambda_j$ matches  the $j$th diagonal element of the sample covariance matrix $S_v = \sum_{i = 1}^n {\boldsymbol v_i^{(0)}} ^\top \boldsymbol v_i^{(0)} / n$, where $\boldsymbol v_{i}^{(0)} = T^{-1}(\boldsymbol x_i^{(0)})$, $\boldsymbol x_i^{(0)}$ is the hyperbolic coordinate from \texttt{hydraPlus}, and $T^{-1}$ is the inverse transformation 
\begin{equation}
  \Tilde{ \boldsymbol{v}} =  (0, \boldsymbol v) = T^{-1}(\boldsymbol x) =  \frac{\operatorname{arccosh}(\alpha)}{\sqrt{\alpha^{2}-1}}(\boldsymbol{x}-\alpha \boldsymbol{\mu}_0^p) \;,
\end{equation}
where $\alpha = -\langle\boldsymbol{\mu}_0^p, \boldsymbol{x}\rangle_{\mathcal{L}}$.

After specifying the prior distributions of $\boldsymbol V, \sigma^2$, and $\Lambda$, the posterior density function of the unknown parameters $\boldsymbol V, \sigma^2$, and $\Lambda$ becomes
\begin{equation}
    \begin{aligned}
p\left(\boldsymbol V, \sigma^{2}, \Lambda \mid D\right) &  \propto  \left(\sigma^{2}\right)^{-(m / 2+a+1)} \prod_{j=1}^{p} \lambda_{j}^{(-n / 2 + \alpha + 1)} \\
\times & \exp \left[-\frac{1}{2 \sigma^{2}} S S R-\sum_{i<j} \log \Phi\left(\frac{\delta_{i j}}{\sigma}\right)-\frac{1}{2} \sum_{i=1}^{n} \boldsymbol v_{i}^\top \Lambda^{-1} \boldsymbol v_{i}-\frac{b}{\sigma^{2}}-\sum_{j=1}^{p} \frac{\beta_{j}}{\lambda_{j}}\right] \; .
\label{eq:post}
\end{aligned}
\end{equation}
Due to the complex form of the posterior density function, we use a Markov-chain Monte Carlo sampler to draw from the posterior distribution.  We describe the sampler in the next section.
\cbend

\section{Posterior Computation}
\label{sec: posterior}

After specifying the Bayesian model and prior choices, we use a Markov-chain Monte Carlo algorithm to sample from the posterior distribution. We first discuss the implementation details of the MCMC algorithm in Section \ref{sec: mcmc}. Then, in Section \ref{sec: case_control}, we present a likelihood approximation based on work of \cite{lscasecontrol} for social networks to accelerate the MCMC with large-scale dissimilarity data. Specifically, we leverage the realization that the posterior distribution can be well approximated using random samples of the objects drawn from a case-control scheme, which reduces the MCMC time complexity from $O(n^2)$ to $O(n)$.

\subsection{Markov-chain Monte Carlo}
\label{sec: mcmc}

\cbstart 
Our MCMC sampler requires the hyperbolic dimension $p$ and the curvature $\kappa$ as inputs.
For hyperbolic curvature, $\kappa$, we can estimate $\widehat \kappa$ using a stress minimization algorithm, which we describe in detail in Appendix \ref{sec: kappa}. For the dimension, $p$, we could use a similar stress minimization approach across potential values of $p$, keeping in mind that the goal is dimension reduction, so our prior is that $p$ is much smaller than $n$.  In practice, we found that a default dimension of $p=2$ works well in a broad range of settings, and it also facilitates interpretation. Another option would be a Bayesian model selection approach similar to the one described by ~\citet{oh2001bayesian}, though we did not pursue that direction in this work. 

Given $p$ and $\kappa$, we initialize the starting values for the MCMC sampler using the output from the \texttt{hydraPlus} algorithm. We take $\boldsymbol V^{(0)} = T^{-1}(\boldsymbol X^{(0)})$ and $\{\delta_{ij}^{(0)}\}_{i, j = 1, \cdots, n}$ as the initial values for $\boldsymbol V$ and $\{\delta_{ij}\}_{i, j = 1, \cdots, n}$. Furthermore, from $\boldsymbol X^{(0)}$, we calculate the initial sum of squared residuals $SSR^{(0)}$ and the sample variance ${\sigma^{2}}^{(0)}=S S R^{(0)} / m$, which can be used as the initial value of $\sigma^{2}$. In addition, the diagonal elements of the sample covariance matrix of $\boldsymbol V^{(0)}$ can be used as initial values for the $\lambda_j$'s.

At each iteration, we first propose a new value of $\lambda_j$ from its conditional posterior distribution given the other unknowns. 
Then, we use a random walk Metropolis-Hasting algorithm to sample the $\boldsymbol v_i$’s and $\sigma^2$. Since we specify a Gaussian prior on $\boldsymbol v_i$, we correspondingly use a normal proposal density. To choose the variance of the normal proposal density, we first write down the full conditional posterior density of $\boldsymbol v_i$, that is,
\begin{equation}
    p\left(\boldsymbol v_{i} \mid \boldsymbol{V}_{-i}, D, \sigma^2, \Lambda, \cdots \right) \propto \exp \left[-\frac{1}{2}\left(Q_{1}+Q_{2}\right)-\sum_{j \neq i, j=1}^{n} \log \Phi\left(\frac{\delta_{i j}}{\sigma}\right)\right] \;,
    \label{eq:condv}
\end{equation}
where $\boldsymbol{V}_{-i} \equiv \{{\boldsymbol{v}_j}: j \neq i, j = 1, \cdots, n\}$, $Q_{1}= \sum_{j \neq i, j=1}^{n}\left(\delta_{i j}-d_{i j}\right)^{2}  / \sigma^2 \text {, and } Q_{2}=\boldsymbol v_{i}^\top \Lambda^{-1} \boldsymbol v_{i}$. Note that, by the model assumption in (\ref{eq:model}), $(\delta_{ij} - d_{ij}) / \sigma$ has a standard normal distribution, approximately, for  $j \neq i, j = 1, \cdots, n$. Thus, $Q_1$ approximately follows a $\chi^2_{n-1}$ distribution. Similarly, since $\boldsymbol{v}_i \sim \mathcal{N}_p(\boldsymbol{0}, \Lambda)$, we have $\boldsymbol v_{i}^\top \Lambda^{-1} \boldsymbol v_{i} \sim \chi^2_p$. Therefore, we have $E(Q_1) \approx n-1 \gg E(Q_2) = p$.
\cbend
Thus, we conclude that $Q_1$ dominates the full conditional posterior distribution, and we may approximate the full conditional variance of $\boldsymbol v_i$ by a constant multiple of $\sigma^2 / (n-1)$, which we use as the variance of the normal proposal density in the MCMC sampler.

Finally, from a preliminary numerical study similar to that carried out by  \cite{oh2001bayesian}, we found that the full conditional density function of $\sigma^2$ can be well approximated by the density function of $IG(m/2 + a, SSR/2 + b)$. 
Moreover, when the number of dissimilarities $m = n(n-1)/2$ is large, the Inverse Gamma density function can be well approximated by a normal density. \cbstart Thus, we approximate the full conditional density of $\sigma^2$ with the Inverse Gamma density to improve computational efficiency, and \cbend use a random walk Metropolis-Hasting algorithm with a normal proposal density and a variance proportional to the variance of $IG(m/2 + a, SSR/2 + b)$ to sample $\sigma^2$.

We now summarize our MCMC algorithm. At iteration $t$: 
\begin{enumerate}
    \item For each $i = 1, \cdots, p$, sample $\lambda_i^{(t)}$ as 
    \begin{equation*}\lambda^{(t)}_i \sim IG\left(\alpha + n / 2, \beta_i + s_i^{(t-1)} / 2\right) \;,
\end{equation*}
where $s_i^{(t-1)}$ is the sample variance of the $i$th coordinates of $\boldsymbol v_{i}^{(t-1)}$'s.
\item For each $i = 1, \cdots, n$, do the following:
\begin{enumerate}
    \item Make a new proposal for $\boldsymbol v_i^{(t)}$ such that 
    \begin{equation*}
    \boldsymbol v_{i, \text{new}}^{(t)} \sim MVN_p\left(\boldsymbol v_i^{(t)}, \frac{c{\sigma^{2}}^{(t-1)}}{n - 1} \cdot I_p\right) \;,
    \end{equation*}
    where $I_p$ is the $p\times p$ identity matrix. In practice, we can simply set the constant multiple $c = 1$. 
    \item Set $\boldsymbol v_i^{(t)} = \boldsymbol v_{i, \text{new}}^{(t)}$ with probability
    \begin{equation*}
        \min\left(1, \frac{p_{\boldsymbol v}(\boldsymbol v_{i, \text{new}}^{(t)})}{p_{\boldsymbol v}(\boldsymbol v_i^{(t)})}\right) \;,
    \end{equation*}
    where $p_{\boldsymbol v}(\cdot) \equiv p(\boldsymbol{v} | \cdots)$ is the full conditional posterior density of $\boldsymbol v$ in (\ref{eq:condv}).
\end{enumerate}
\item Make a new proposal for ${\sigma^{2}}^{(t)}$ such that 
\begin{equation*}
    {\sigma^{2}}^{(t)}_{\text{new}} \sim N\left( {\sigma^{2}}^{(t)}, \frac{c \gamma^{(t)}}{(\omega-1)^{2}(\omega-2)}\right) \;,
\end{equation*}
where ${\gamma}^{(t)} = (SSR^{(t)} / 2 + b)^2$ and $\omega = m/2 + a$. Set ${\sigma^{2}}^{(t)}={\sigma^{2}}^{(t)}_{\text{new}}$ with probability
\begin{equation*}
    \min\left(1, \frac{p_{\sigma^{2}}({\sigma^{2}}^{(t)}_{\text{new}})}{p_{\sigma^{2}}({\sigma^{2}}^{(t)})}\right) \;,
\end{equation*}
where $p_{\sigma^{2}}(\cdot)$ is the density function of $IG(m/2 + a, SSR^{(t)}/2 + b)$ which approximates the full conditional density of $\sigma^2$.
\end{enumerate}

\cbstart
Using the above algorithm, we obtain samples from the full posterior density. We use the posterior median $\widehat{\sigma}^2$ of the MCMC samples as our estimate of the measurement error. The estimates of $\{\delta_{ij}\}$ and $\boldsymbol{X}$ are more involved. Each $\{\delta_{ij}\}$ corresponds to a unique set of $\boldsymbol{X}$ up to Procrustean operations, so estimating $\{\delta_{ij}\}$ is equivalent to estimating $\boldsymbol{X}$. Moreover, we cannot directly use sample medians to estimate $\delta_{ij}$, as there is no guarantee that there exists a certain set of hyperbolic coordinates the distances among which are the sample medians $\{\delta_{ij}\}$. Instead, we use the posterior sample of $\{\delta_{ij}\}$ that minimizes the SSR as our estimate of the posterior mode. We observe that the likelihood in (\ref{eq:lik_new}) dominates the posterior density in (\ref{eq:post}), and the term involving the SSR in (\ref{eq:lik_new})
 dominates the likelihood, so the posterior mode of $\{\delta_{ij}\}$ should be well approximated by the posterior sample of $\{\delta_{ij}\}$ that minimizes the SSR. Moreover, the value of  $\{\delta_{ij}\}$ that minimizes the SSR also minimizes the stress, as the goodness-of-fit measure stress is just the square root of SSR after normalization. 
 
 We then define the Bayesian estimate of $\{\delta_{ij}\}$ as the posterior stress-minimizing estimate 
\begin{equation}
    \{\widehat \delta_{ij}\} \equiv \arg \min_{\{\delta_{ij}\}^{(t)}} SSR^{(t)} = \arg \min_{\{\delta_{ij}\}^{(t)}} \text{stress}^{(t)} \;,
    \label{eq:mode}
\end{equation}
where the superscript $(t)$ indicates that the posterior sample is drawn from the $t$-th MCMC iteration, and take the Bayesian estimate of $\boldsymbol X$ as $\boldsymbol X^{(t)}$. 
In general, $\boldsymbol{X}$ is invariant to isometric actions (rotation, reflection, etc.) on $\mathbb{H}^p(\kappa)$
  and is thus not identifiable. We can only recover the relative embedding of the objects instead of their absolute hyperbolic coordinates. \cite{oh2001bayesian} suggested post-processing the MCMC samples of $\boldsymbol X$ at each iteration of the MCMC via the Procrustes operation, so that the transformed $\boldsymbol X^\prime$ has sample mean $0$ and a diagonal covariance matrix as specified in the prior. We found in  simulations, however, that the MCMC algorithm mixed well without post-processing and returned accurate estimates of the model parameters. Therefore, we skipped this post-processing step, which has the added benefit of speeding up our algorithm, since the Procrustes transformation involves an eigen-decomposition of a large matrix.

  \cbend

\subsection{Case-control Likelihood Approximation}
\label{sec: case_control}

For dissimilarity data with a sample size of around $n < 200$, $20,000$ iterations of the MCMC algorithm take about $300$ seconds using a standard personal computer. However, since the proposal of $\boldsymbol v_i$ involves calculations across $n$ terms, and for each iteration, we need to update $\boldsymbol v_i$ $n$ times in total, the time complexity of each iteration is approximately $O(n^2)$. When the sample size $n$ increases, the algorithm quickly becomes computationally intractable. 

We propose a stratified case-control log-likelihood to approximate the original logarithm of posterior likelihood to facilitate computation using larger datasets.  Our case-control approach is based on work by \cite{lscasecontrol} for social networks, where they studied a posterior log-likelihood approximation of the latent space model described in \cite{Hoff}. The core intuition is that, for each object $i$ fixed, if we stratify all other objects regarding their dissimilarities to $i$, then there are many fewer objects that are similar to $i$ than there are objects that are very dissimilar. This imbalance creates an opportunity to borrow ideas from the widely-used case-control technique from epidemiology. The samples in a case-control study can also be stratified into two distinct groups, where the ``case" group has the outcome of interest, but is often rare and hard to collect compared to the ``control" group. 

This suggests that the statistical approximation technique in case-control studies can be used to approximate the posterior distribution of $\boldsymbol v_i$. If we view the objects similar to object $i$ as samples in the case group, and the rest as being in the control group, we can approximate the posterior distribution using all the samples in the case group and a random sample from the control group. Moreover, to increase precision, we further stratify the samples in the control group by their dissimilarities to object $i$, and randomly sample from each stratum by their weight determined by their contributions to the proposal likelihood change to enhance the accuracy of the approximation. Under the proposed case-control stratification scheme, this reduces the MCMC time complexity from $O(n^2)$ to $O(n)$.

We will now give details of the stratified case-control log-likelihood. The logarithm of the full conditional posterior density of $\boldsymbol v_i$ is
\begin{equation}
    l_i \equiv \log \pi\left(\boldsymbol v_{i} \mid \cdots\right) \propto -\sum_{j \neq i, j = 1}^n \left[\frac{(\delta_{ij} - d_{ij})^2}{2\sigma^2} + \log \Phi\left(\frac{\delta_{i j}}{\sigma}\right)\right] - \frac{1}{2}\boldsymbol v_i ^\top \Lambda^{-1} \boldsymbol v_i\;.
\end{equation}
The first step is to divide object $j = 1, 2, \cdots, n, j \neq i$ into $M$ different strata $S_1^{(i)}, S_2^{(i)}, \cdots, S_M^{(i)}$ according to their observed dissimilarities with respect to object $i$. The partition of the strata can be highly customized, as long as the total number of strata $M \ll n$. We will later describe several partition strategies in Section \ref{sec:case_ex} and \ref{sec: gene}.
Given the strata, we can write the log-likelihood as 
\begin{equation}
\begin{aligned}
l_i = -\sum_{k = 1}^M \sum_{j \in S_{k}^{(i)}}\left[\frac{(\delta_{ij} - d_{ij})^2}{2\sigma^2} + \log \Phi\left(\frac{\delta_{i j}}{\sigma}\right)\right] - \frac{1}{2}\boldsymbol v_i ^\top \Lambda^{-1} \boldsymbol v_i = \sum_{k = 1}^M l_{i, k} - \frac{1}{2}\boldsymbol v_i ^\top \Lambda^{-1} \boldsymbol v_i \;,
\label{eq: strata}
\end{aligned}
\end{equation}
where $l_{i, k}= -\sum_{j \in S_{k}^{(i)}}\left[(\delta_{ij} - d_{ij})^2 / 2\sigma^2+ \log \Phi\left(\delta_{i j} / \sigma\right)\right]$ is the likelihood contribution from stratum $S_{k}^{(i)}$.

If a stratum $S_{k}^{(i)}$ belongs in the case group, we will compute its log-likelihood explicitly. Otherwise, if $S_{k}^{(i)}$ belongs in the control group, we will randomly sample $n_{i, k}$ objects from the stratum $ S_{k}^{(i)}$ and estimate the stratum's log-likelihood contribution by 
\begin{equation}
    \hat{l}_{i, k} = -\frac{N_{i, k}}{n_{i, k}} \sum_{j \in n_{i,k}\text{ samples}} \left[\frac{(\delta_{ij} - d_{ij})^2}{2\sigma^2} + \log \Phi\left(\frac{\delta_{i j}}{\sigma}\right)\right] \;,
    \label{eq: est_log}
\end{equation}
where $N_{i, k}$ is the number of elements in stratum $S_{k}^{(i)}$. Since the estimator $\widehat{l}_{i, k}$ is based on a random sample from the strata, we always have $\mathbb{E}(\widehat{l}_{i, k}) = {l}_{i, k}$, so that the log-likelihood estimator is unbiased. For the sake of analysis, we now assume the first $C$ strata are considered as cases, which we denote by $S_{1}^{(i)}, S_{2}^{(i)}, \cdots, S_{C}^{(i)}$, and the rest are controls. Then, the stratified case-control approximate log-likelihood for object $i$ becomes 
\begin{equation}
\begin{aligned}
        & \hat{l}_i = \sum_{k = 1}^C l_{i, k} + \sum_{k = C+1}^M \hat{l}_{i, k} - \frac{1}{2}\boldsymbol v_i ^\top \Lambda^{-1} \boldsymbol v_i\;,
    \label{eq:case}
\end{aligned}
\end{equation}
where $l_{i,k}$ is the $S_k^{(i)}$'s likelihood contribution in (\ref{eq: strata}) and $\hat{l}_{i,k}$ are the log-likelihood estimators in (\ref{eq: est_log}).

We now describe how to determine the subsample size $n_{i, k}$. To accelerate the MCMC iteration to approximately $O(n)$, we want object $i$'s overall random sample size $n_i \equiv \sum_{k = 1}^M n_{i, k} \ll n$. To do this, we pick a moderate control-to-case rate $r$, and let $n_i$ be $r$ times the average number of objects in the case group. That is, we set
$n_i \equiv r \cdot \frac{1}{n} \sum_{i = 1}^n \sum_{k = 1}^C {|S_{k}^{(i)}|}$, where ${|S_{k}^{(i)}|}$ denotes the number of objects in $S_{k}^{(i)}$. Given the fixed $n_i$, we assign $n_{i, k}$ proportional to the stratum  $ S_{k}^{(i)}$'s contribution to the log-likelihood change in sampling $\boldsymbol v_i$. 

We conduct the following pilot MCMC to determine the stratum  $S_k^{(i)}$'s likelihood contributions. For object $i$ fixed, we first draw a simple random sample over its control group of size $n_i$, and 
use them to construct an approximate log-likelihood
\begin{equation}
    \Tilde{l}_{i} \equiv \sum_{k = 1}^C l_{i, k}
    + \frac{n - \sum_{k = 1}^C{|S_{k}^{(i)}|}}{n_i} \sum_{j \in n_{i}\text{ samples}} \left[\frac{(\delta_{ij} - d_{ij})^2}{2\sigma^2} + \log \Phi\left(\frac{\delta_{i j}}{\sigma}\right)\right] - \frac{1}{2}\boldsymbol v_i ^\top \Lambda^{-1} \boldsymbol v_i \;,
    \label{eq:pilot}
\end{equation}
which we will use in the pilot MCMC. Once again, since we randomly sample from the population, $\Tilde{l}$ is unbiased. Then, at each iteration $t$, we calculate the log-likelihood change for $\boldsymbol v_i$ as 
\begin{equation*}
    \begin{aligned}
    \Delta \Tilde l_i^{(t)} &= \Tilde{l}_i (\boldsymbol v_{i, \text{new}}^{(t)}) -\Tilde{l}_i (\boldsymbol v_{i}^{(t)}) \\
    & =\sum_{k = 1}^C \left[l_{i, k} (\boldsymbol v_{i, \text{new}}^{(t)}) - l_{i, k} (\boldsymbol v_{i}^{(t)})\right] + \sum_{k = C + 1}^M\left[\Tilde{l}_{i, k} (\boldsymbol v_{i, \text{new}}^{(t)}) - \Tilde{l}_{i, k} (\boldsymbol v_{i}^{(t)})\right] + \Delta_i \\
    & = \sum_{k = 1}^C \Delta l_{i, k} + \sum_{k = C + 1}^M \Delta \Tilde{l}_{i, k} + \Delta_i \;,
    \end{aligned}
\end{equation*}
where $\Delta_i = -(\boldsymbol v_{i, \text{new}} ^\top \Lambda^{-1} \boldsymbol v_{i, \text{new}} -  \boldsymbol v_{i} ^\top \Lambda^{-1} \boldsymbol v_{i}) / 2$. We then define 
\begin{equation*}
    w^{(t)}_{i,k} = \left|\frac{\Delta \Tilde{l}_{i, k}}{\sum_{g = C+ 1}^M\Delta \Tilde{l}_{i, g}}\right| \quad \text{for }k = C+1, C+2, \cdots, M \;,
\end{equation*}
and calculate the relative weights as
\begin{equation*}
    w_{i,k} = \frac{1}{T - 1} \sum_{t = 1}^{T-1} w^{(t)}_{i,k} \;,
\end{equation*}
where $T$ is the number of iterations in the pilot MCMC run after burn-in and thinning. Finally, we take the subsample size $n_{i, k}$ as
\begin{equation*}
    n_{i, k} = n_i \cdot \frac{w_{i, k}}{\sum_{g = c+ 1}^M w_{i, g}} \;,
\end{equation*}
for $k = c+1, \cdots, M$. 

To summarize, the algorithm is as follows:
\begin{enumerate}
    \item For each object $i = 1, \cdots, n$, partition all other objects into $M$ strata via a user-defined, dissimilarity-based strategy.
    \item Set the strata defined with small dissimilarities $S_{1}^{(i)}, S_{2}^{(i)}, \cdots, S_{C}^{(i)}$ as cases, and the rest as controls.
    \item For each object $i$, randomly sample $n_i$ objects from the control group.
    \item Run a pilot MCMC with the approximate log-likelihood described in (\ref{eq:pilot}).
    \item Record the relative weights $w_{i, k}$ and compute the subsample sizes $n_{i, k}$.
    \item For each object $i = 1, 2, \cdots, n$, sample $n_{i, k}$ objects from strata $S_{k}^{(i)}$ for $k = C+1, C+2, \cdots, M$.
    \item Run a full MCMC with the original log-likelihood functions $l_1, l_2, \cdots, l_n$ replaced by the stratified case control log-likelihood functions in (\ref{eq:case}).
\end{enumerate}

In the following sections, we evaluate both full and approximate strategies for sampling from the posterior using both simulated and observed data. 

\section{Simulation Experiments}
\label{sec: simulations}

We conducted simulation experiments to evaluate aspects of the proposed statistical model and algorithm. First, we evaluate BHMDS's element-wise estimation performance for the true dissimilarities, $\delta_{ij}$, and the measurement error variance, $\sigma^2$. Then, we evaluate the overall estimation performance using the coverage rate of the posterior credible interval (CI) over all the $\{\delta_{ij}\}$. Lastly, we evaluate the robustness of the algorithm by examining its calibration under a variety of data generating distributions.

To evaluate BHMDS's estimation performance, we wish to test it under different dataset sizes $n$, hyperbolic dimensions $p$, and noise levels $\sigma$. Throughout our experiments, we fix the hyperbolic curvature $\kappa = 1$, and generate the simulation data as follows. 
\begin{enumerate}
\cbstart
    \item For each combination of $(n, p, \sigma) \in \{50, 100\} \times \{2, 5\} \times \{1, 1.5, 2\}$, sample $\boldsymbol X = \{\boldsymbol x_1, \cdots, \boldsymbol x_n\}$ from the hyperbolic wrapped normal distribution such that $\boldsymbol x_1, \cdots, \boldsymbol x_n \sim_{i.i.d.} \mathcal{G}\left( \boldsymbol 0_p, 3I_p\right)$, where $\mathcal{G}(\cdot, \cdot)$ is the hyperbolic wrapped normal distribution centered at the origin and $I_p$ is the $p$-dimensional identity matrix. Specifically, the wrapped normal distribution $\mathcal{G}(\cdot, \cdot)$ samples from a normal distribution on the tangent space at the hyperbolic origin in $\mathbb{R}^p$, projects the tangent space onto the hyperbolic space by the transportation described in (\ref{eq:map}), results in a Gaussian-like distribution on $\mathbb{H}^p$ and generates tree-like data.
    \cbend
    \item Then, for each set of $(n, p, \sigma)$, generate 50 sets of noisy dissimilarity matrices $\{d_{ij}\}_{i, j = 1,\cdots, n}$, with entries $d_{ij}$ drawn from (\ref{eq:model}).
    \item Apply the full BHMDS MCMC to the $\{d_{ij}\}$'s and record the approximate posterior mode estimates $\{\widehat{\delta}_{ij}\}$ described in (\ref{eq:mode}), posterior mean $\widehat{\sigma}$, and the matrix-wise converge rate
    \begin{equation}
    C \equiv \frac{\sum_{i < j} I\left(\delta_{ij} \in \left[q_{ij}^{(\alpha / 2)}, q^{(1 - \alpha / 2)}_{ij}\right]\right)}{m} \;,
    \label{eq: ci}
\end{equation}
where $\left(q_{ij}^{(\alpha / 2)}, q^{(1 - \alpha / 2)}_{ij}\right)$ are the $(\alpha / 2)$ and $(1 - \alpha /2)$ quantiles of the posterior samples of ${\delta}_{ij}$ and $m = n(n-1) / 2$.
\end{enumerate}

\begin{figure}
    \centering
    \includegraphics[width = 0.6\linewidth]{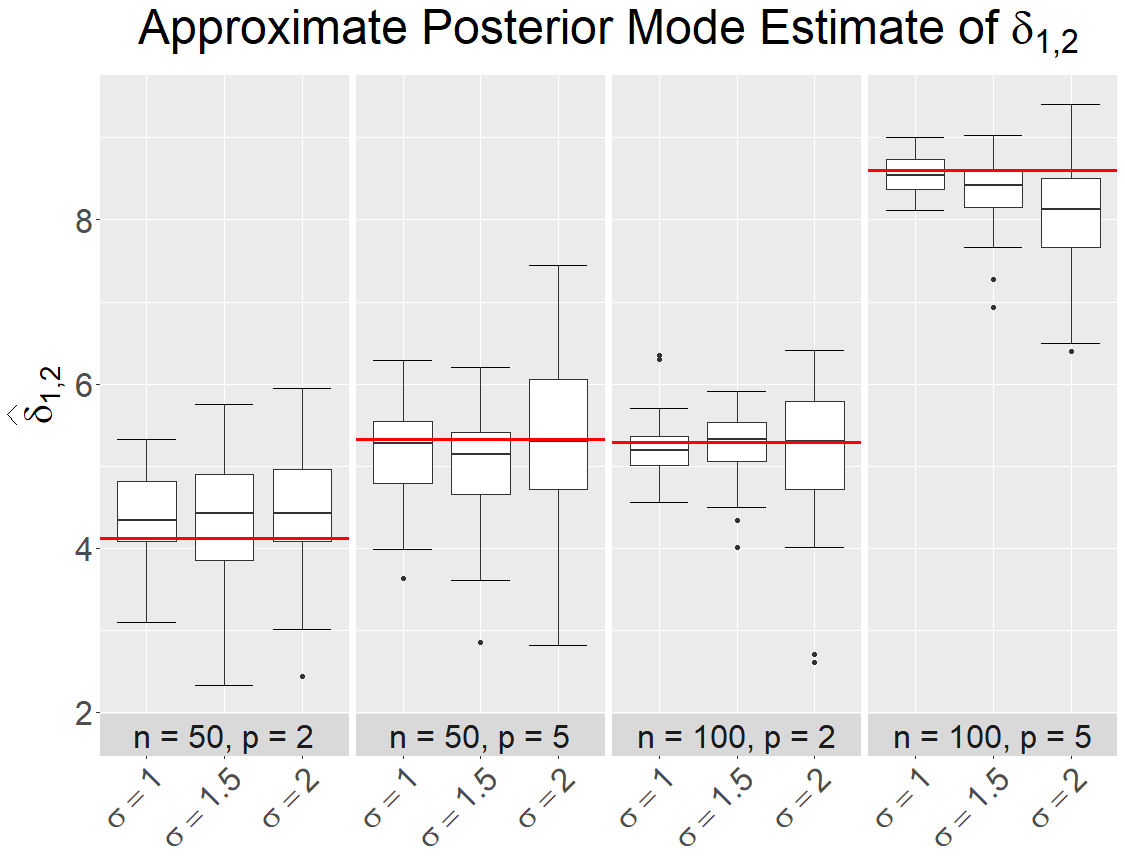}
    \caption{\footnotesize{Simulation results of the BHMDS estimation performance on the true dissimilarity $\delta_{1, 2}$. 
    The facet labels, i.e., $n = 50$, $p = 2$, correspond to the sample size $n$ and hyperbolic dimension $p$ of the true dissimilarity data, with $(n, p) \in \{50, 100\} \times \{2, 5\}$. The x-axis labels, e.g. $\sigma = 1$, correspond to the noise level of the observed dissimilarity data, with $\sigma \in \{1, 1.5, 2\}$. Without loss of generality, we summarized the results on $\widehat{\delta}_{1,2}$, the approximate posterior mode estimate of the dissimilarity between objects 1 and 2, in the box plots above. Each box plot corresponds to 50 estimates of $\widehat{\delta}_{1,2}$ at a level of $(n, p, \sigma)$, and the red lines in each facet correspond to the true dissimilarity measures $\delta_{1, 2}$ at level $(n, p)$. All the box plots are concentrated around the true values, indicating that BHMDS precisely and robustly predicts the true dissimilarity measure.}}
    \label{fig: post_delta}
\end{figure}

\begin{figure}
    \centering
    \includegraphics[width = 0.6\linewidth]{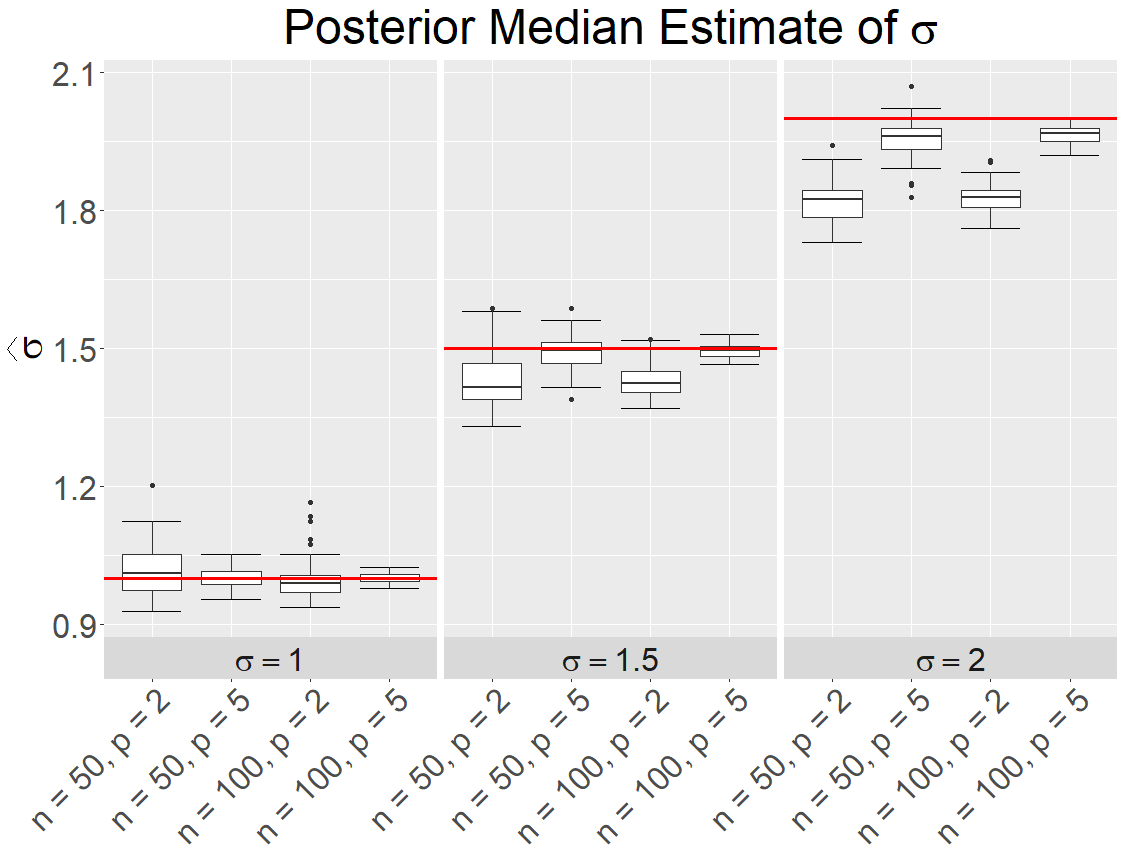}
    \caption{\footnotesize{Simulation result of the BHMDS estimation performance on the true measurement error $\sigma$. 
    The facet labels, i.e. $\sigma = 1$, correspond to the noise level of the observed dissimilarity data, with $\sigma \in \{1, 1.5, 2\}$. The x-axis labels, i.e., $n = 50$, $p = 2$, correspond to the sample size $n$ and hyperbolic dimension $p$ of the true dissimilarity data, with $(n, p) \in \{50, 100\} \times \{2, 5\}$.  We summarized the results on $\widehat{\sigma}$, the posterior median estimate of $\sigma$, in the box plots above. Each box plot corresponds to 50 estimates of $\widehat{\sigma}$ at a level of $(n, p, \sigma)$, and red lines in each facet correspond to the true noise level $\sigma$ value. We conclude that, although the accuracy degenerates with large observational errors, BHMDS performs reasonably well in estimating $\sigma^2$. Moreover, we observe that BHMDS is more robust to errors with larger hyperbolic dimensions.}}
    \label{fig: post_sigma}
\end{figure}

\begin{figure}
    \centering
    \includegraphics[width = 0.6\linewidth]{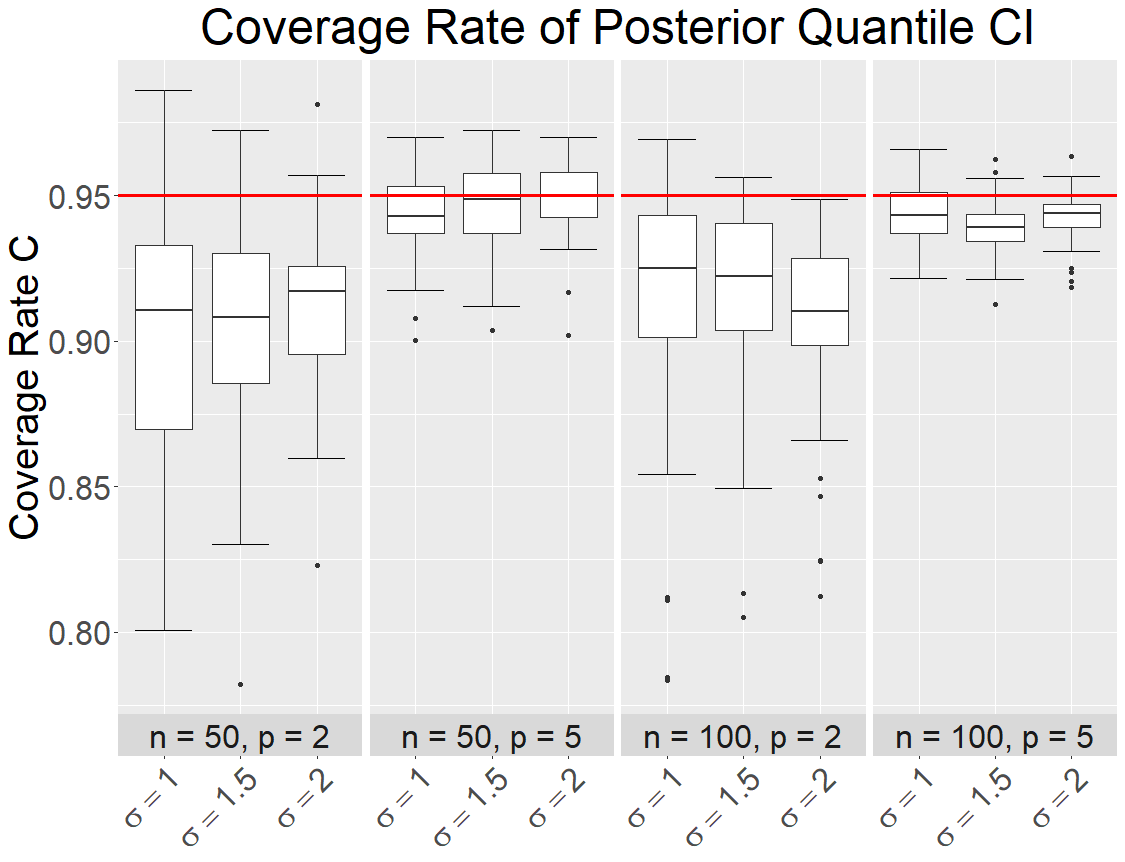}
    \caption{\footnotesize{Simulation result of the BHMDS credible interval's coverage performance.
    The facet labels, i.e., $n = 50$, $p = 2$, correspond to the sample size $n$ and hyperbolic dimension $p$ of the true dissimilarity data, with $(n, p) \in \{50, 100\} \times \{2, 5\}$. The x-axis labels, i.e. $\sigma = 1$, correspond to the noise level of the observed dissimilarity data, with $\sigma \in \{1, 1.5, 2\}$. Each box plot contains $50$ matrix-wise coverage rates corresponding to the 50 noisy observations, calculated as described in (\ref{eq: ci}). The red lines in each facet correspond to the nominal $95\%$ coverage rate. We can observe that BHMDS achieves close-to-nominal coverage rates at all levels of $(n, p, \sigma)$, and the coverage improves as the sample size $n$ and hyperbolic dimension $p$ increases.}}
    \label{fig: CI}
\end{figure}

We plot the simulation results for $\{\widehat{\delta}_{ij}\}$, $\widehat{\sigma}$, and coverage rate in Figures \ref{fig: post_delta}, \ref{fig: post_sigma}, and \ref{fig: CI} respectively. For any combination of $(n, p, \sigma)$, all the boxplots are concentrated tightly around the red horizontal lines representing the true values in all three plots, indicating that BHMDS yields precise and robust estimates of $\delta_{ij}$ and $\sigma^2$.
\cbstart
\cbend

\cbstart
\cbend

\cbstart
\cbend

\cbstart
We are interested in further testing the robustness of the BHMDS algorithm using dissimilarity data generated via different distributions defined with a variety of dimensions, curvatures, and
distribution parameters. To this end, we introduce marginal calibration, a criterion that comprehensively assesses the predictive performance of the forecasting distribution described in \cite{calib}. For simplicity, we only give the intuition here, and one may  refer to \cite{calib} for technical details. Suppose at times or instances $s = 1, 2, \cdots, S$, nature picks distributions $G_1, G_2, \cdots, G_S$, and we predict them with forecasting distributions $F_1, F_2, \cdots, F_S$ calculated from a certain forecasting mechanism. Marginal calibration theory shows, if the forecasting mechanism predicts the quantities of interest well, the empirical distribution of $x_1, x_2, \cdots, x_S$, which are observations randomly sampled from $G_1, G_2, \cdots, G_S$ at time $1, 2, \cdots, S$, should resemble the average of the predictive distribution. That is,
\begin{equation}
\widehat{G}_{S}(x) \equiv \frac{1}{S} \sum_{s=1}^{S} \mathbf{1}\left(x_{s} \leq x\right) \approx \widehat{F}_{S}(x) \equiv \frac{1}{S} \sum_{s = 1}^S F_s(x)
 \quad 
\text{for all } x \;.
\end{equation}
Thus, if we plot the difference $\widehat{F}_{S}(x) - \widehat{G}_{S}(x)$ for all $x$, and if the forecasting mechanism is robust, we should observe the difference fluctuating in a close neighborhood of zero. 
\cbend

To evaluate the robustness of BHMDS via marginal calibration, we need to first specify the distributions  $\{G_s\}_{s = 1, 2, \cdots, S}$ and  $\{F_s\}_{s = 1, 2, \cdots, S}$. We cannot choose the wrapped normal distribution $\mathcal{G}_s(\boldsymbol x)$ as $G_s$, since $G_s$ is required to be univariate, yet $\mathcal{G}_s(\boldsymbol x)$ is a function of the multivariate random variable $\boldsymbol x$, which we wish to vary under different hyperbolic dimensions. Thus alternatively, we choose the distribution of $\delta_{ij}$ as $G_s$, which is the distribution of the hyperbolic distance between coordinates $\boldsymbol x_i$ and $\boldsymbol x_j$ randomly sampled from a given $\mathcal{G}_s(\boldsymbol x)$, and choose the $F_s$ as the distribution of $\widehat{\delta}_{ij}$ estimated by BHMDS. Choosing such a $G_s$ is beneficial, as it is inherently univariate and is uniquely determined by the data generating distribution $\mathcal{G}_s(\boldsymbol x)$, so that if BHMDS estimates the true dissimilarity well, the distribution of $\widehat{\delta}_{ij}$ will resemble the distribution of $\delta_{ij}$. Given $G_s$'s and $F_s$'s, for each $s = 1, 2, \cdots, 1000$, we randomly generate the parameters 
\begin{align*}
    &p^{(s)} \sim \text{Multinomial}(2, 3, 4, 5) \text{ with equal probability}\;,\\
        & \kappa^{(s)} \sim \text{Unif}(0.2, 2) \;, \\
    &\mu_p^{(s)} \sim \mathcal {N}_p (0, 2I_p) \;, \\
    &\Sigma_{ii}^{(s)} \sim_{i.i.d.} \text{Unif}(5, 10) \text{, and } \Sigma_{ij}^{(s)} = 0 \text{ for all } i, j = 1, 2, \cdots, p, i \neq j \;.
\end{align*}
For each set of parameters, we simulate 20 sets of noisy dissimilarity matrices at noise level $\sigma = 1$. 
We then apply BHMDS to the noisy matrices and record the estimated dissimilarities $\{\widehat\delta_{ij}\}^{(s)}_1, \{\widehat\delta_{ij}\}^{(s)}_2, \cdots, \{\widehat\delta_{ij}\}^{(s)}_{20}$, $i, j = 1, 2, \cdots, n$.

\begin{figure}
    \centering
    \includegraphics[width = 0.6\linewidth]{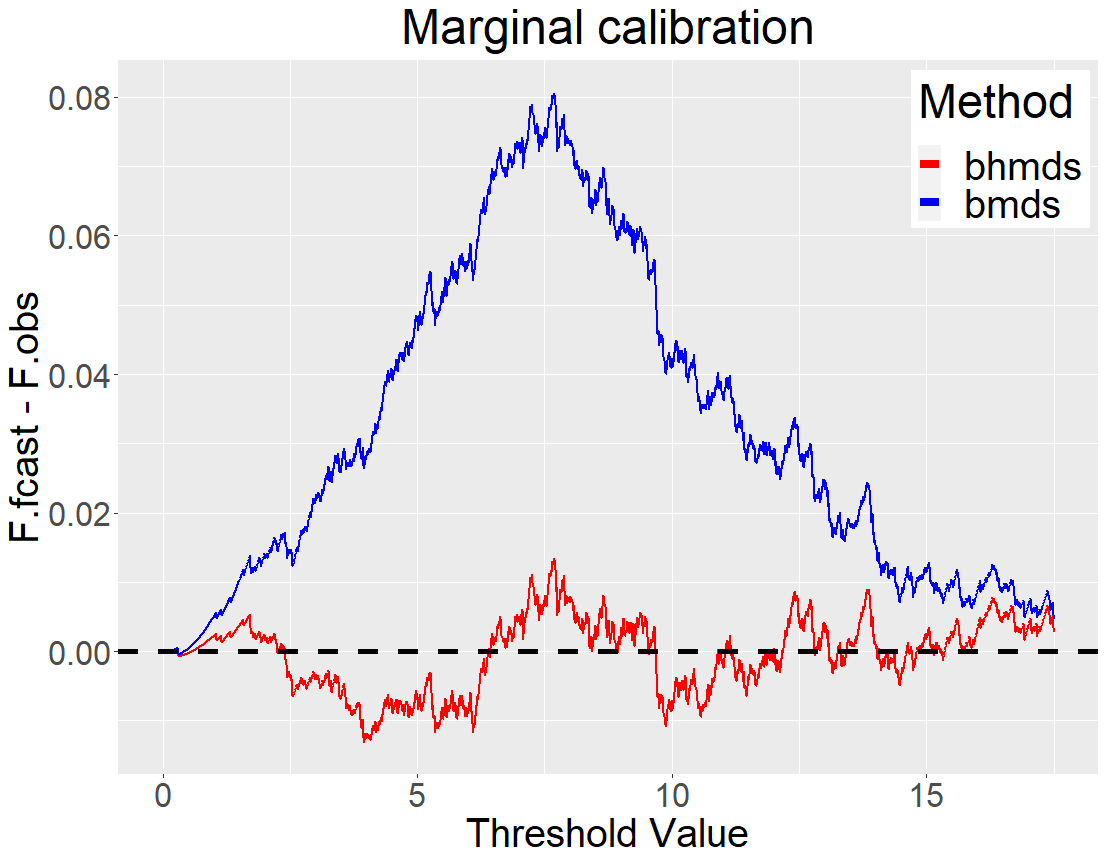}
    \caption{\footnotesize{The marginal calibration result for BHMDS amd BMDS. The x-axis corresponds to the threshold values of the dissimilarity distribution. The y-axis corresponds to the difference $\widehat{F}_{S}(x) - \widehat{G}_{S}(x)$. The red line corresponds to the marginal calibration plot of BHMDS, the blue line corresponds to the marginal calibration plot of BMDS, and the black horizontal dashed line corresponds to $y = 0$.  We can observe that when the true dissimilarities are generated from the hyperbolic geometry, BHMDS performs significantly better than BMDS, with the difference only fluctuating in a small interval around zero, in contrast to the BMDS difference, where there is a large spike at $x = 7.5$. This suggests that BHMDS outperforms BMDS in estimating dissimilarities when the data are hierarchical.}}
    \label{fig: calib}
\end{figure}

Given the distributions $G_s$ at each instance $s$, we can sample $x_1, x_2, \cdots, x_S$ explicitly and use them to construct $\widehat{G}_S(x)$. On the other hand, we cannot directly access the $F_s$'s, but we can estimate them by the empirical CDFs $\Tilde{F}_s$'s from the posterior estimates. To minimize the correlation between the samples, we use the samples $\{\widehat\delta_{i, i + 1}\}^{(s)}_{1}, \{\widehat\delta_{i, i + 1}\}^{(s)}_{2}, \cdots, \{\widehat\delta_{i, i + 1}\}^{(s)}_{20}$ for $i = 1, 2, \cdots, n-1$ to compute $\Tilde{F}_s$'s. Finally, we plot the difference in Figure \ref{fig: calib} in red. We further evaluate the marginal calibration of the Euclidean BMDS via the same process and plot the difference in blue. The BHMDS method is much better calibrated than the BMDS method, as the calibration curve for the BHMDS method is closer to zero for all the threshold values on the $x$-axis. This suggests that when the dissimilarity data are hierarchical, tree-like, or have an intrinsic hyperbolic property, the proposed BHMDS algorithm yields much more precise estimates of the true dissimilarity compared to Euclidean MDS method. 

\section{Data Analysis}
\label{sec: data_analysis}

We now use the proposed BHMDS algorithm to analyze several dissimilarity datasets. First, we compare BHMDS with extant MDS algorithms using MDS goodness-of-fit criteria with respect to several hierarchical dissimilarity datasets collected in social network and Natural Language Processing (NLP) studies. Then we present a case study of the case-control likelihood approximation on hierarchical NLP hypernym data. Lastly, we apply BHMDS with global human gene expression data to investigate the cellular differentiation of different cell types with confidence quantification on their rank statistics. 

\subsection{Comparison with Existing MDS Approaches}

In this section, we apply the BHMDS algorithm to several tree-like, hierarchical datasets and evaluate its embedding performance via MDS goodness-of-fit criteria stress and distortion. 

We first give the definitions of the MDS goodness-of-fit criteria. Stress is one of the most prevalent criteria in the MDS literature, which measures the $L_2$ goodness-of-fit of the embedding. Given the observed dissimilarities $\{d_{ij}\}$ and its MDS embedding $\{\widehat{\delta}_{ij}\}$, their stress is defined as
\begin{equation}
    \text{stress}\left(\{d_{ij}\}, \{\widehat{\delta}_{ij}\}\right) \equiv \sqrt{\frac{\sum_{i<j}\left(d_{i j}-\widehat{\delta}_{i j}\right)^{2}}{\sum_{i<j} d_{i j}^{2}}}\; ,\quad i, j = 1, \cdots, n\;.
\end{equation}The distortion, on the other hand, measures the $L_1$ goodness-of-fit of the embedding, defined as
\begin{equation}
    \text{Distortion}\left(\{d_{ij}\}, \{\widehat{\delta}_{ij}\}\right) \equiv \frac{1}{m} \sum_{i < j} \frac{\left|d_{ij} - \widehat\delta_{ij}\right|}{d_{ij}}\; ,\quad i, j = 1, \cdots, n \;,
\end{equation}
where $m = n(n-1)/2$ is the number of dissimilarities. Both criteria normalize the difference terms by the observed dissimilarities, which enables cross-comparison among datasets with different sizes and scales.

We consider the following datasets in our numerical study. We first consider the Zachary's Karate Club dataset, a commonly used social network of a university karate club, first described by \cite{karate} and studied via the hyperbolic MDS algorithms \texttt{hydra} and \texttt{hydraPlus} in \cite{hydra}. Next,  we consider a phylogenetic tree dataset that expresses the genetic heritage of mosses growing in urban environments, described by \cite{moss} and studied in \cite{tradeoff}. We further consider two tree-like, hierarchical datasets: one is a Computer Science Ph.D. advisor-advisee network, available from \cite{csphd} and also studied in \cite{tradeoff}; the other is the WordNet mammals subtree dataset, an NLP hypernym dataset studied in \cite{poinword}.
All of the four datasets come in the form of undirected graphs, thus we take the shortest path lengths on the graph between objects $i$ and $j$ as the observed dissimilarity measures $\{d_{ij}\}$.

\cbstart Given the discrete nature of the shortest path lengths, the Gaussian error structure may not be the optimal choice. Yet from numerical experiments and later applications, we find that the residuals fitted from the shortest path lengths closely follow a Gaussian distribution. We therefore conclude that the Gaussian error model works well with shortest path lengths as dissimilarities. \cbend Given the observed dissimilarity matrices, we fix the hyperbolic dimension as $p = 2$ and estimate their curvature $\kappa$'s as described in Appendix \ref{sec: kappa}, except for the karate dataset, where we set $\kappa = 1$ as in \cite{hydra}. Given the hyperbolic curvature and dimension, we use our BHMDS algorithm to compute the posterior estimate $\{\widehat \delta_{ij}\}$ and the corresponding goodness-of-fit criteria. To compare the embedding performance with existing methods, we included results from common hyperbolic MDS methods using the same hyperbolic curvature and dimension, as well as Euclidean MDS methods with dimension $p = 2$. 

The embedding results are displayed in Table \ref{tab: stress} and \ref{tab: dist}. We can conclude from the tables that the BHMDS algorithm attains optimal or close-to-optimal embedding in terms of both criteria on all four datasets. This indicates that, when the dissimilarity data are tree-like or hierarchical, the proposed BHMDS algorithm not only quantifies the uncertainty in the observed dissimilarity data but also embeds them into hyperbolic geometry with minimal information loss compared to the state-of-the-art MDS algorithms.

\begin{table} 
\caption{\footnotesize{Embedding performance of the four datasets in terms of the stress criteria. The red text corresponds to the optimal stress values, and the blue text corresponds to the second best value. We observe that the stress-minimizing hyperbolic MDS methods, namely BHMDS and \texttt{hydraPlus}, constantly outperform all other methods, and their embedding results are comparable. This indicates that tree-like, hierarchical data are best represented on hyperbolic geometry in terms of stress. Moreover, it shows that the BHMDS algorithm can be used to optimize the minimal-stress embedding.}}
\vspace{20pt}
\centering
\renewcommand{\arraystretch}{0.8}
 \begin{tabular}{c c c c c c c c c} 
 \hline
& Size $n$ & BHMDS& \texttt{hydraPlus}& \texttt{hydra}& BMDS& \texttt{smacof}& \texttt{cmds}\\
 \hline
Karate $(\kappa = 1)$ & 34 &
\color{blue} 0.1780 &
\color{red}0.1727 &
0.2105 &
0.2050 &
0.2112 &
0.2850 \\
Phylo $(\kappa = 0.14)$ &344 &
\color{blue} 0.0413 &
\color{red}0.0413 &
0.2091 &
0.1601 &
0.1594 &
0.2481  \\
CS phd $(\kappa = 0.55)$ & 1025 & 
\color{red}0.1469 &
\color{blue}0.1471 &
0.2029 &
0.2281 &
0.2351 &
0.3862\\
Wordnet $(\kappa = 2.06)$ & 1141&
\color{red}0.0798 &
\color{blue}0.0807 &
0.1279 &
0.2722 &
0.2725 &
0.4928\\
 \hline
 \end{tabular}
 \vspace{10pt}
 \label{tab: stress}
\end{table}

\begin{table} 
\caption{\footnotesize{Embedding performance of the four datasets in terms of the Distortion criterion. The red text corresponds to the optimal Distortion values, and the blue text corresponds to the second best value. Again, we can observe that, the stress-minimizing hyperbolic MDS methods, namely BHMDS and \texttt{hydraPlus}, outperform the other methods even though they are not designed to optimize the Distortion, and their embedding results are comparable. This indicates that tree-like, hierarchical data are also best represented on hyperbolic geometry in terms of Distortion. Moreover, it shows that the BHMDS algorithm can be used to optimize the minimal-distortion embedding.}}
\vspace{20pt}
\centering
\renewcommand{\arraystretch}{0.8}
 \begin{tabular}{c c c c c c c c c} 
 \hline
&Size $n$ & BHMDS& \texttt{hydraPlus}& \texttt{hydra}& BMDS& \texttt{smacof}& \texttt{cmds}\\
 \hline
Karate $(\kappa = 1)$ & 34& 
\color{blue} 0.3485 &
\color{red}0.3287 &
0.4278 &
0.3986 &
0.3742 &
0.5234\\
Phylo $(\kappa = 0.14)$ &344 &
\color{blue} 0.1214 &
\color{red}0.1206 &
0.6806 &
0.3552 &
0.3516 &
0.5918\\
CS phd $(\kappa = 0.55)$ &1025& 
\color{blue} 0.2869 &
\color{red}0.2868 &
0.4545 &
0.4523 &
0.4510 &
0.7513\\
Wordnet $(\kappa = 2.06)$ &1141&
\color{red}0.1441 &
\color{blue} 0.1460 &
0.2242 &
0.4994 &
0.4967 &
0.9670\\
 \hline
 \end{tabular}
 \vspace{10pt}
 \label{tab: dist}
\end{table}

\subsection{Approximated Log-likelihood: A Case Study}
\label{sec:case_ex}

In this section, we present a case study to exemplify how to apply the stratified case-control log-likelihood and evaluate the algorithm's likelihood precision and computational efficiency. We consider the WordNet mammals subtree dataset, a hierarchical NLP hypernym dataset studied by \cite{poinword}, as the input dataset. The WordNet dataset is in the form of an undirected graph with $n = 1141$ nodes, and we take the shortest path lengths on the graph between objects $i$ and $j$ as the observed dissimilarities $d_{ij}$. We set the hyperbolic dimension $p = 2$ and estimate the curvature using the methods described in Appendix \ref{sec: kappa}.

We now explain how to apply the approximated log-likelihood to the WordNet dataset. For each object $i$, we first partition objects $j \in \{1, 2, \cdots, n,  j \neq i\}$ by their observed dissimilarity $d_{ij}$'s. Since we are using the shortest path length as dissimilarity, all $d_{ij}$'s are positive integers ranging from $1, 2, \cdots, \max_j (d_{ij})$ for each $i$, and objects with the same $d_{ij}$ value are in the same group. That is, we set 
\begin{equation}
    S_k^{(i)} \equiv \left\{\text{Object } j: d_{ij} = k, j = 1, 2, \cdots, n, j \neq i\right\}, \quad k = 1, 2,\cdots, \max_j(d_{ij}) \;,
\end{equation}
so that we collect all objects of distance $k$ to object $i$ in $S_{k}^{(i)}$. 

We consider $S^{(i)}_1$ and $S^{(i)}_2$, the two strata defined with the smallest dissimilarities (neighbors and second-neighbors) as cases and all other $S_{j}^{(i)}$ for $j > 2$ as controls. We choose a control-to-case rate $r = 5$ so that in the approximated log-likelihood, we explicitly calculate over $7\%$ of the terms in the exact likelihood. We run a pilot MCMC with 3,000 iterations, of which the first 1,000 are discarded as burn-in, to compute the relative weights and sample sizes for each $S^{(i)}_{j}$ for $j > 2$. We then run the case-control approximated log-likelihood MCMC to obtain the posterior estimate $\{\delta_{ij}\}$.

We first evaluate the overall estimation performance of the case-control MCMC algorithm in terms of the stress and computation time. We run both the approximate and full MCMC algorithm with the WordNet dataset, and record the stress value of their posterior estimates as well as the computational times per 100 MCMC iterations. The results are shown  in Table \ref{tab: case}. We observe that the case-control approximate log-likelihood MCMC achieves comparable stress to the full MCMC but with only half of the computation time, indicating that the approximate MCMC achieves fast and accurate estimates of the true dissimilarities for large datasets.

\begin{figure}
    \centering
    \includegraphics[width = 0.6\linewidth]{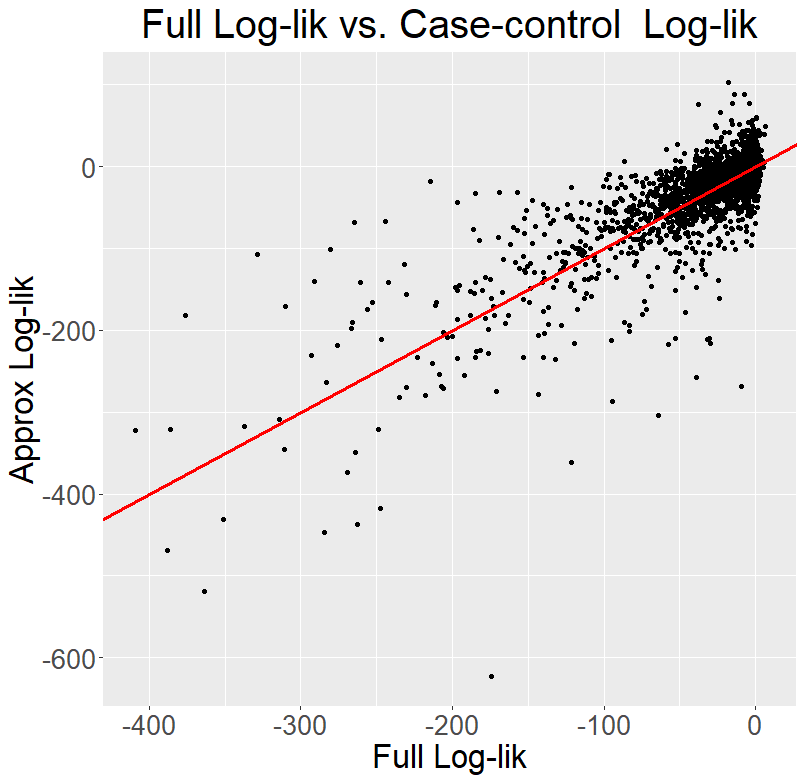}
    \caption{\footnotesize{X-axis: Log-likelihood change values calculated via the full MCMC algorithm. Y-axis: Log-likelihood values calculated via the case-control approximate MCMC algorithm. The red line corresponds to the line $y = x$. We observe that, the approximated log-likelihood changes are tightly aligned around $y = x$ against the full log-likelihood changes, with a strongly positive correlation $\rho = 0.82$. This indicates the log-likelihood values computed from the proposed case-control MCMC algorithm are a good approximation of the true log-likelihood values, and the case-control MCMC shares a similar accept/reject pattern as the full MCMC.}}
    \label{fig: case}
\end{figure}

To evaluate the precision of the case-control likelihood, we compare the case-control log-likelihood change to the full log-likelihood change in the MCMC proposal of $\boldsymbol v_i$.  If the case-control log-likelihood change approximates the full log-likelihood change well, the case-control MCMC will accept or reject in a similar pattern as the full MCMC. Recall that the change in the log-likelihood, as defined in Section \ref{sec: case_control}, is
$\Delta \Tilde l_i^{(t)} = \Tilde{l}_i (\boldsymbol v_{i, \text{new}}^{(t)}) -\Tilde{l}_i (\boldsymbol v_{i}^{(t)})$.
For a valid comparison, it is essential to evaluate both likelihood changes with respect to the same $\boldsymbol v_i$ proposal and parameters such as $\{\delta_{ij}\}, \sigma^2, \Lambda, \beta$. To this end, we first run the full MCMC algorithm with the Wordnet dataset, and record 100 sets of parameters $\boldsymbol \theta^{(i)} = \left(\{\delta_{ij}\}^{(i)}, (\sigma^2)^{(i)}, \Lambda^{(i)}, \beta^{(i)}\right)$ from 100 MCMC iterations. For computational efficiency, we randomly sample 100 objects from the $n = 1141$ samples for each $\boldsymbol \theta^{(i)}$ and make a proposal upon the corresponding $\boldsymbol v_1^{(i)}, \boldsymbol v_2^{(i)}, \cdots, \boldsymbol v_{100}^{(i)}$ under $\boldsymbol \theta^{(i)}$. We compute the approximate likelihood change and the full likelihood change for each proposal $\boldsymbol v_j^{(i)}, i,j = 1, \cdots, 100$, and plot the approximated log-likelihood change values against the full log-likelihood change values in Figure \ref{fig: case}. We observe that the approximate log-likelihood change is tightly aligned around $y = x$, with a strongly positive correlation $\rho = 0.82$. This indicates the log-likelihood values computed from the proposed case-control MCMC algorithm are a good approximation to the true log-likelihood values. Furthermore, the case-control MCMC shares a similar accept/reject pattern as the full MCMC and is able to properly sample from the posterior.

\begin{table} 
\caption{\footnotesize{Stress values and computational time per 100 MCMC iterations of the WordNet mammal subtree dataset with \texttt{hydraPlus}, approximated MCMC, and full MCMC. We can observe that, the approximated case-control algorithm achieves a comparable stress with \texttt{hydraPlus} and the full MCMC, indicating it estimates a close-to-optimal hyperbolic embedding of the WordNet dataset. More importantly, the proposed case-control algorithm is approximately twice as fast as the full MCMC algorithm. This will enable the BHMDS framework scalable with dissimilarity data of large sample sizes.}}
\vspace{20pt}
    \centering
    \begin{tabular}{c c c c}
    \hline
        Method & hydraPlus & Approx & Full \\
        \hline
         Stress & 0.081 & 0.085 & 0.080 \\
         MCMC time (100 iters) & - & 28.39s & 54.88s \\
         \hline
    \end{tabular}
    \vspace{10pt}
    \label{tab: case}
\end{table}
\cbstart
\subsection{Assessing Network Centrality With BHMDS}
\label{sec: network}

We now use BHMDS to identify high-centrality nodes in a social network. Network centrality measures how centrally a node or actor is positioned in the network \citep{centrality}. Various centrality measures have been developed to describe the relative importance of a node in the network, such as degree centrality, closeness centrality, betweenness centrality, and eigenvector centrality \citep{centralindice}. Centrality is useful in social and economic analysis. Among others, \cite{leadercloseness} shows that recognized leaders will most likely emerge at positions with a high closeness centrality of closeness. \cite{eigenvectordiffusion} show that actors with a high eigenvector centrality facilitate the spread of information. 

In this example, we focus on closeness centrality, which is defined as the inverse of the average shortest path length from a given node to all other nodes in the network. By definition, the higher the closeness centrality, more proximate a node is to all other nodes on average. We claim that when using the shortest path lengths as dissimilarities to embed networks on hyperbolic space, the distance from the origin to a node's embedded hyperbolic coordinates resembles the node's closeness centrality. Similar centrality results are also reported in \cite{hyperclose2} and \cite{hyperclose1}. Nodes that exhibit relatively equal and short distances to all other nodes can be interpreted as being located in the higher hierarchy of the network, so that by the nature of the hyperbolic geometry, it will be more likely to be embedded around the hyperbolic origin. Similarly, nodes that are relatively far from all the other nodes are more likely to be embedded far from the origin. This characteristic enables hyperbolic MDS methods to effectively assess network closeness.

\begin{figure}
    \centering
    \includegraphics[width = \linewidth]{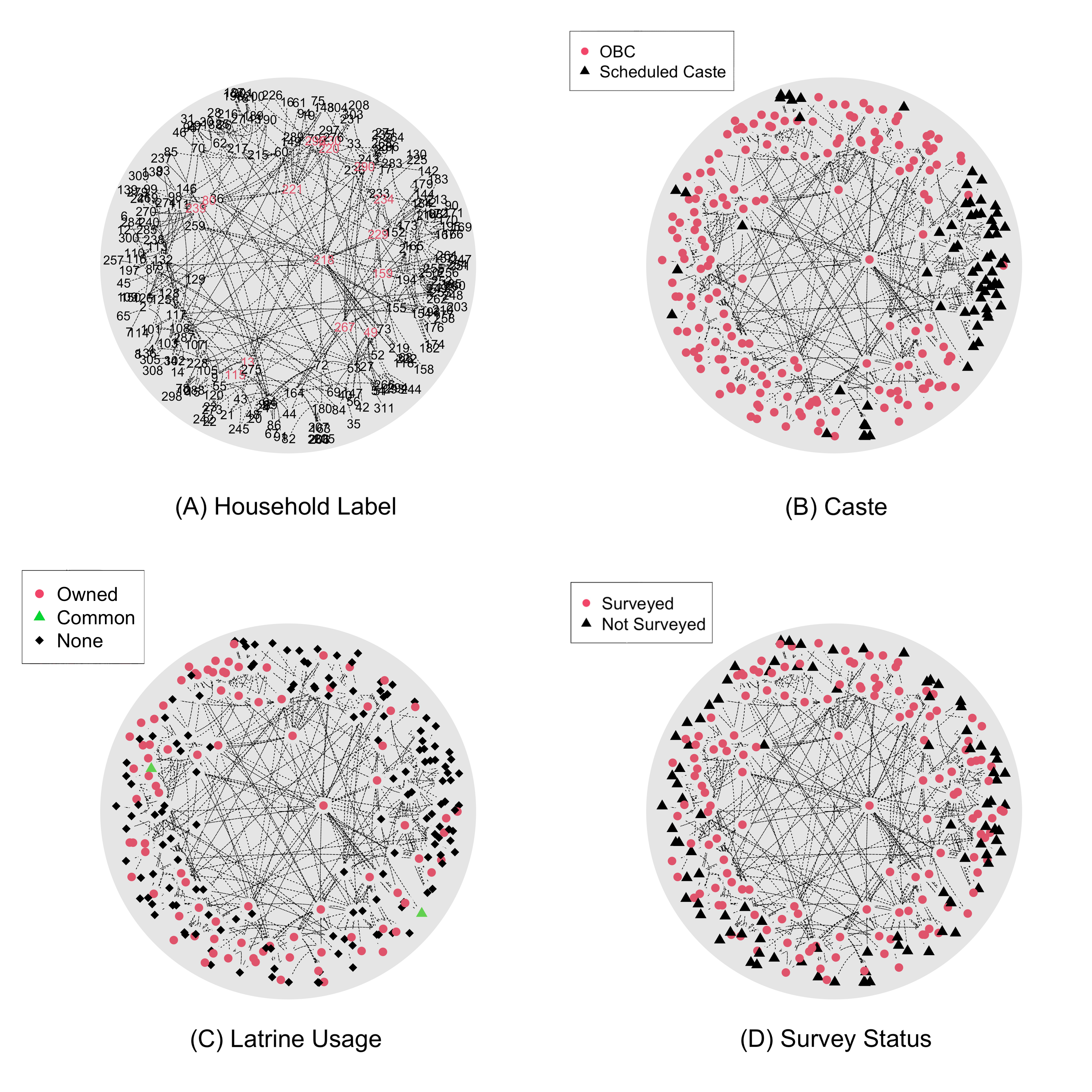}
    \caption{\footnotesize{(A) Visualization of the BHMDS embedded result with household label. The red labels correspond to the 15 closest households recognized by BHMDS. (B): Visualization by caste type: Other backward class (OBC) or Scheduled caste. (C) Visualization by household latrine usage: Owned a latrine, Use a common latrine, or Do not own a latrine. (D) Visualization by if the household is surveyed.}}
    \label{fig: networkvisual}
\end{figure}

\begin{figure}
    \centering
    \includegraphics[width = 0.6\linewidth]{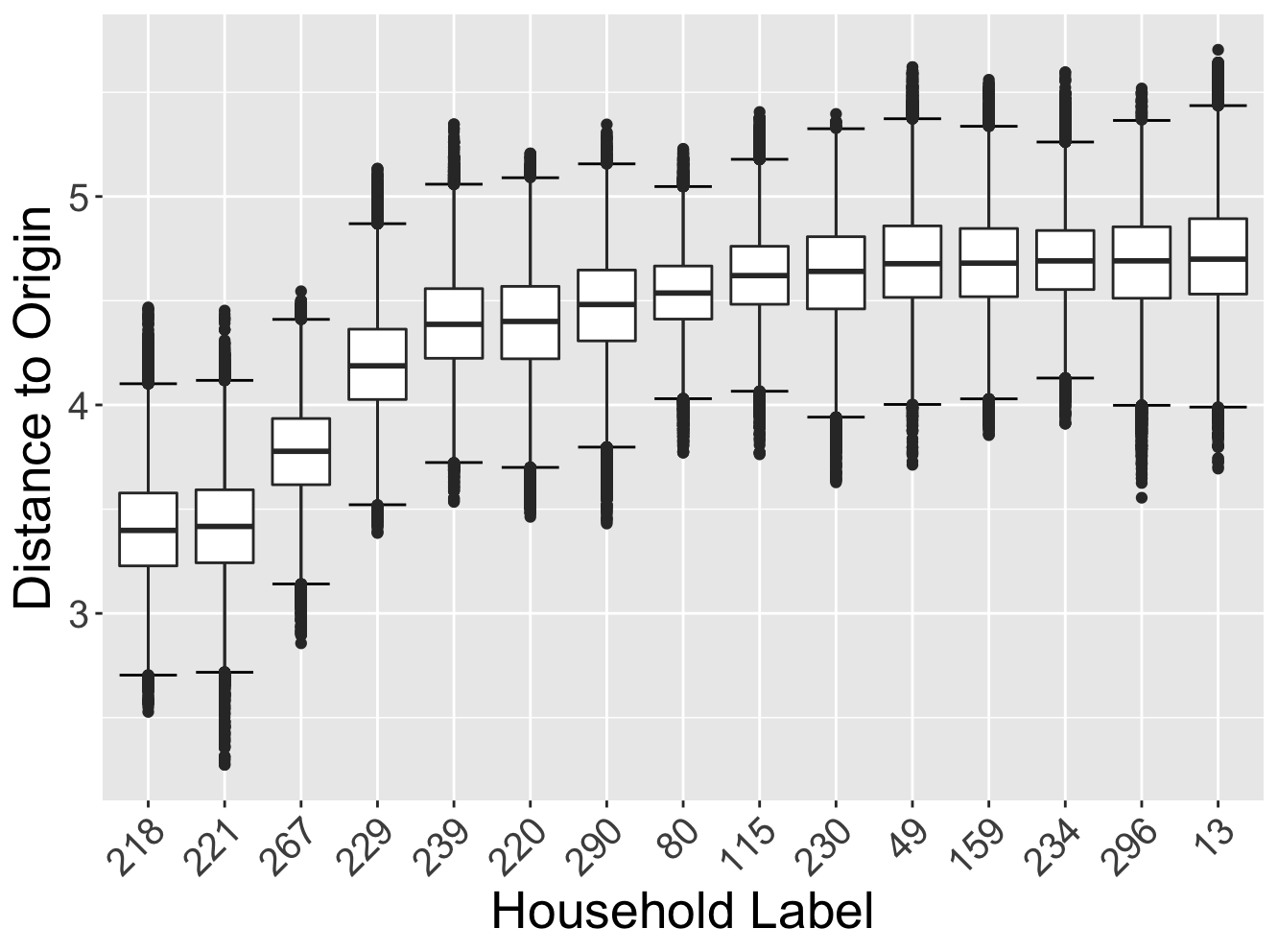}
    \caption{\footnotesize{MCMC samples of the coordinate-to-origin distances of the 15 closest household recognized by BHMDS. We observe that the closest five households (218, 221, 267, 229, and 239) also have the largest closeness centrality in the network. Households 218, 221, and 267 are distinctively closer to the origin, which indicates they have higher centrality. For the rest of the households, although there exists a gradient in the sample medians, the highly overlapped sample boxplots suggest they should be treated homogeneously.}}
    \label{fig: networkdist}
\end{figure}

We apply BHMDS to financial networks from villages in rural India. The connections in these networks represent borrowing behaviors. Our goal is to access network closeness with uncertainty quantification and visualization. The network is collected by \cite{eigenvectordiffusion} and describes the relationship of borrowing money between 248 households in a rural area of southern Karnataka, India. A household census, along with demographic covariates such as caste, religion, latrine usage, survey status, etc., is also collected. We use a public release of these data that contains a sample of households and can be found in \url{https://dataverse.harvard.edu/dataset.xhtml?persistentId=doi:10.7910/DVN/U3BIHX}. We first visualize the BHMDS embedding on the Poincaré disk in Figure \ref{fig: networkvisual} to identify potential nodes with high closeness and to observe the relations between covariates. For an optimal visualization layout, we first set $\kappa = 1$. From (A), we observe that Households 218, 221, and 267 are among the closest to the origin, which also has the largest closeness centrality in the network. From (B) and (C), we observe a distinct cluster of households of scheduled caste on the right side of the disk, which systematically do not own a latrine. Moreover, in (D), we observe that households that do not participate in the survey appear at the edge of the disk. The ties between the unsurveyed households are mostly recorded from the surveyed households, and the ties between the unsurveyed households are missing. 

To further quantify the uncertainty in network centrality, we first run BHMDS repeatedly over a curvature interval and find that the stress is minimized with $\kappa = 9.6$. Using this curvature, we run BHMDS again and record all MCMC samples of each node's distance to the hyperbolic origin. Here, the optimal curvature $\kappa$ is large, which potentially complicates the likelihood surface and makes optimization algorithms such as \texttt{hydraPlus} unstable. Specifically, it usually takes \texttt{hydraPlus} more than 60,000 iterations to converge, and households closest to the origin recognized by \texttt{hydraPlus} are not consistent across different trials. In contrast, even though using inconsistent \texttt{hydraPlus} as starting values, BHMDS consistently embeds the households with the highest closeness centrality the closest to the origin and achieves better stress than \texttt{hydraPlus}. Furthermore, in Figure \ref{fig: networkdist} we observe that Households 218, 221, and 267 are distinctly closer to the origin. For all other nodes, although there is a gradient in the medians, the boxplots of their MCMC samples overlap significantly. Such uncertainty in the samples suggests that the centrality of these households is close and should be considered homogeneously. Together, these results suggest that BHMDS has some advantages over \texttt{hydraPlus} for this dataset.
\cbend
\subsection{Quantifying Uncertainty in Human Gene Expression Data}
\label{sec: gene}

We now use the proposed BHMDS algorithm to analyze a global human gene expression dataset. \cite{lukk} integrated microarray data from 5372 human samples representing 369 different cells \& tissue types, disease states, and cell lines, and constructed a global human gene expression map. The original data come in the form of a gene expression matrix of over 22000 probe sets times 5372 genes of 15 human cell types.
\cbstart
The expression matrix is jointly normalized by sample and library size, so that we can directly construct the dissimilarity matrix using the pairwise Euclidean distance of the gene vectors.
\cbend
Although the dissimilarities are computed from a Euclidean embedding,  \cite{sharpee} proposed that the intrinsic geometry of the human gene expression data is hyperbolic. Moreover,  \cite{noisegene1}, \cite{noisegene2}, and \cite{noisegene3} argued that gene expression data are likely to contain measurement error. Together, these considerations motivate us to apply BHMDS with the human gene expression dissimilarity matrix to quantify the uncertainty in the data. Specifically, we use the case-control approximate MCMC algorithm to compute the Bayesian estimate of the cluster distances between cell type communities and quantify the uncertainty on the rank statistics of cell types' evolution pseudotimes. 

We first elaborate on the details of the implementation of the case-control approximated MCMC. As the dissimilarity matrix is computed from high dimensional Euclidean metrics, the original human gene expression dissimilarities are large, which leads to overflow issues in computation. Thus, we preprocess the dissimilarity data as follows. We fix the hyperbolic curvature to $\kappa = 1$, scale the dissimilarity matrix by a constant, and use \texttt{hydraPlus} to compute the corresponding stress with the hyperbolic dimension $p = 5$ as chosen in \cite{sharpee}. We repeat the above process with a grid search over the scaling constants until we find an optimal constant. Such a process is similar in spirit to the algorithm described in Section \ref{sec: kappa}, as altering the curvature is roughly equivalent to scaling the distance as observed in (\ref{eq: dist}), with the relationships between the dissimilarities unchanged. We found that scaling the dissimilarity by 10 yields the optimal stress at 0.046, resulting in a rescaled matrix with an average dissimilarity at 15.12. Since the dissimilarity in gene expression is continuous, we partition the rescaled dissimilarity by continuous intervals $[0, 7),[7, 8),[8, 9), \cdots, [24, \infty)$, and consider the dissimilarities that fall within $[0,7)$ as the case group. We choose $r = 10$, so that we will explicitly compute $6\%$ of the likelihood terms. 

\begin{figure}
    \centering
    \includegraphics[width = \linewidth]{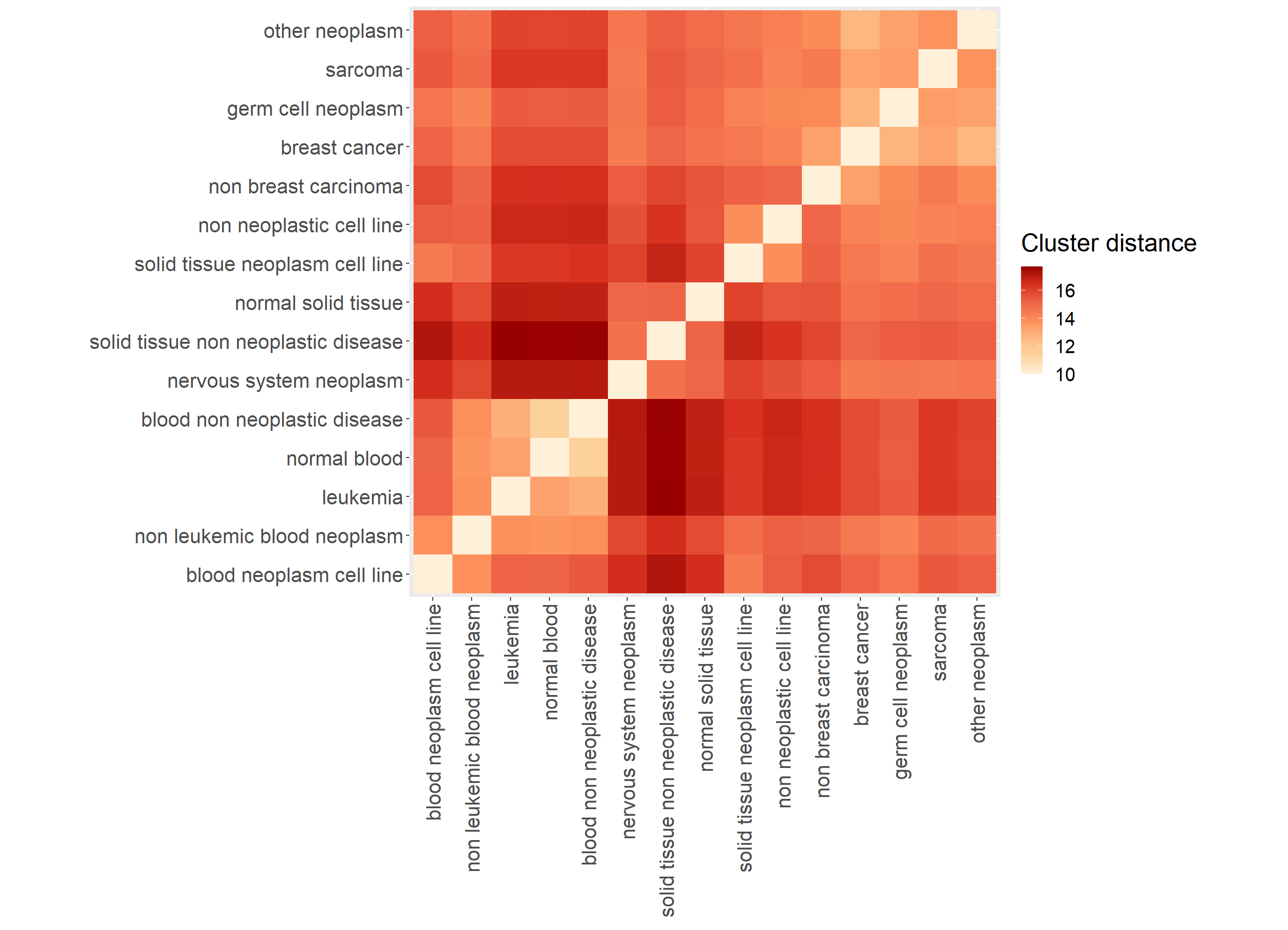}
    \caption{\footnotesize{Heatmap of the cluster distance between cell type communities defined in (\ref{eq: d_cluster}). We observe the hematopoietic cell types, i.e., blood neoplasm cell line, non-leukemic blood neoplasm, leukemia, normal blood, and blood non-neoplastic disease, are distinct from all other cells. We also observed modularity in neoplasm cells, i.e. the clustering of breast cancer cell, germ cell neoplasm, sarcoma, and other neoplasm cells.}}
    \label{fig: heatmap_2}
\end{figure}

To compute the Bayesian estimate of the cell type cluster distance, at each MCMC iteration, we record the cluster distance for cell type community $C_i$ and $C_j$ of size $n_i$ and $n_j$ as \begin{equation}
\label{eq: d_cluster}
    d_{\text{cluster}}(C_i, C_j) = \frac{1}{n_i n_j} \sum_{(k, l) \in (C_i, C_j)} d_{\mathbb{H}} (\boldsymbol x_k, \boldsymbol x_l), \quad i \neq j, \quad i, j = 1, 2, \cdots, 15\;,
\end{equation}
where the community membership is given in the original data. We then take the posterior median of each $d_{\text{cluster}}(C_i, C_j), i, j = 1, \cdots, 15$, and plot them in Figure \ref{fig: heatmap_2}. 
Our Bayesian estimate identifies  hematopoietic cells as distinct from all other cells, which is also observed in \cite{lukk}. We also observe modularity in neoplasm cells in the upper right corner of Heatmap \ref{fig: heatmap_2}, indicating their proximity in terms of evolutionary distance.

We further used the BHMDS algorithm to measure different cell types' cellular differentiation using rank statistics with uncertainty quantification, a feature that is not available from previous methods used to analyze these data. Cellular differentiation refers to the transition of immature cells into specialized types, which is a central task in modern developmental biology \citep{pseudotime}. Specifically, \cite{pseudotime} studied the hierarchy of the single-cell data on hyperbolic geometry. 
For cells that are less differentiated, or at the beginning of a developmental process, they will be at the root of the evolutionary hierarchy and have relatively equal and small distances from all other cells, so that by the nature of the hyperbolic geometry, they are more likely to be embedded around the hyperbolic origin. Similarly, for cells that are more differentiated, they will have relatively large distances from all other cells, so they are more likely to be embedded distant from the origin. Thus, \cite{pseudotime} proposed that the hyperbolic distance between the origin and the cell's embedded coordinate can be a good measure of the extent of the cell's cellular differentiation, which they denoted as hyperbolic evolution pseudotime. 

\begin{figure}
    \centering
    \includegraphics[width =\linewidth]{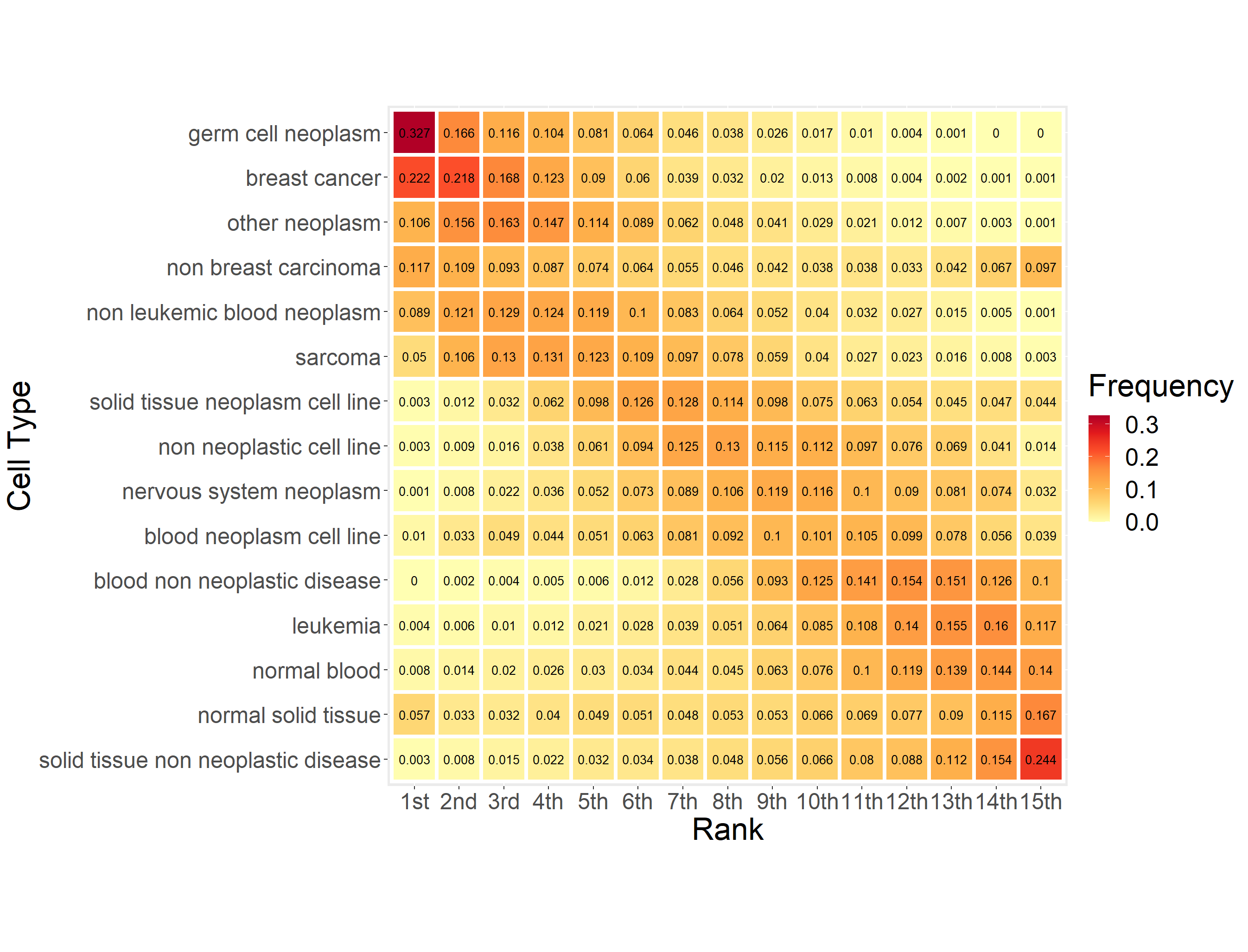}
    \caption{\footnotesize{Heatmap on the frequency of the rank statistics for 15 different human cell types. We observe that cell types within the same cluster (hematopoietic, neoplasm) share similar hierarchies, with the neoplasm cell on the higher hierarchy and the hematopoietic cells on the lower hierarchy. The germ cell neoplasm most frequently attains the smallest evolution pseudotime, entailed by breast cancer cells and other neoplasm cells. This potentially indicates that the neoplasm cells with high frequency in rank statistics are more de-differentiated.}}
    \label{fig: heatmap_1}
\end{figure}

In our study, at each MCMC iteration, we record the sampled evolution pseudotime for each gene and use them to construct the posterior credible intervals of the evolution pseudotime for each gene. To visualize the average evolution pseudotime for each cell type, we summarized the rank statistics of each cell type as follows. For each cell type community, we randomly sample one of its genes and draw from its nominal pseudotime confidence interval based on the posterior. Then, we rank the pseudotime drawn from each cell type community and record their rank statistics. We repeat the above process 10,000 times. We then summarize the rank statistics by frequency in Figure \ref{fig: heatmap_1}, where the $ij$-th entry of the heatmap represents the frequency of cell type $i$ being the $j$-th closest to the hyperbolic origin, which indicates it is the $j$-th least differentiated. 

Figure \ref{fig: heatmap_1} indicates that cell types within the same cluster in Figure \ref{fig: heatmap_2} share a similar hierarchy, as the neoplasm cells obtain higher frequency for the higher rank statistics and thus are less differentiated, whereas the hematopoietic cells yield higher frequency for the lower rank statistics, and thus are more differentiated. \cbstart Additionally, we observe that the germ cell neoplasm has the smallest evolution pseudotime, entailed by the breast cancer cell. These results further shed light on the study of cancer stem cells, as short evolution pseudotimes of the two neoplasm cells suggest they are likely to be less differentiated or more de-differentiated, a phenomenon often observed in malignant tumors. \cbend

\section{Conclusion}
\label{sec: conclusion}
We have proposed a Bayesian approach to multidimensional scaling in hyperbolic space. Using a previously studied generating process for observed dissimilarities from \cite{oh2001bayesian}, we used prior distributions on hyperbolic space to derive the posterior distribution of the model parameters. We then proposed an MCMC procedure to sample from this posterior and also proposed a quick case-control method to efficiently sample from the posterior when the size of the observed dissimilarity matrix is large. Finally, we applied our methods to datasets in several domains and showed how our Bayesian procedure leads to embeddings with low distortion and allows us to quantify the uncertainty due to noise in the observed dissimilarities. 

\cbstart
The proposed method uses a state-of-the-art method called \texttt{hydraPlus} to choose parameters of the prior distribution and initialize the MCMC sampler. The major distinction between BHMDS and \texttt{hydraPlus} is that BHMDS quantifies the uncertainty in the dissimilarity data, which often leads to analyses that are more precise. Specifically, in Section \ref{sec: network}, we show that BHMDS outperformed \texttt{hydraPlus} in some example datasets in identifying social network nodes with high closeness centrality using the extra uncertainty assessment. Through numerical experiments, we also find BHMDS often achieves better stress and gives stable estimates than \texttt{hydraPlus} when the sample size, hyperbolic curvature, or measurement error is large. Together this suggests that BHMDS numerically outperforms hydraPlus while adding to uncertainty quantification. 
\cbend

In independent work issued just as we submitted this paper,~\citet{Praturu2022.10.12.511940} also carried out a Bayesian analysis of hyperbolic MDS. The present paper builds on~\citet{liu_lubold_raftery_mccormick_2022}. The present work differs from that of~\citet{Praturu2022.10.12.511940} in several ways.  Namely, we use priors for the latent positions in the hyperbolic space whereas~\citet{Praturu2022.10.12.511940} assume a Gaussian noise structure on distances. We choose to use priors on the positions as we consider a hyperbolic wrapped normal prior that allows us to reparameterize the posterior likelihood and greatly facilitate the MCMC sampler.  Additionally, we propose using a case-control approximation to address computation whereas~\citet{Praturu2022.10.12.511940} use an iterative approach that adds observations in blocks.

There are several potential avenues of future research. First, we have assumed in this work a low dimension for the desired embedding. While this does allow for easy visualization of the resulting embedding, imposing a low dimension on the dissimilarities is likely to lead to higher distortion than higher dimensions. One potential area of future work could derive an information criterion to estimate an optimal embedding dimension given the dissimilarities, as was done in \cite{oh2001bayesian}. A second possible direction is to explore how our proposed procedure could be adopted into the non-metric multidimensional scaling framework. Finally, while hyperbolic space has received considerable attention in the past few years, spaces of non-constant curvature, as studied in \cite{tradeoff} and others, might lead to an embedding with lower distortions. An interesting research question here could focus on proposing Bayesian MDS methods in these spaces.

\bibliographystyle{abbrevnamed.bst}
\bibliography{bhmds}

\begin{thebibliography}{}

\bibitem[\protect\citeauthoryear{Banerjee \bgroup \em et al.\egroup
  }{2013}]{eigenvectordiffusion}
A.~Banerjee, A.~G. Chandrasekhar, E.~Duflo, and M.~O. Jackson.
\newblock The diffusion of microfinance.
\newblock {\em Science}, 341(6144):1236498, 2013.

\bibitem[\protect\citeauthoryear{Bavelas}{1950}]{leadercloseness}
A.~Bavelas.
\newblock Communication patterns in task‐oriented groups.
\newblock {\em Journal of the Acoustical Society of America}, 22:725--730, 11
  1950.

\bibitem[\protect\citeauthoryear{Borg and Groenen}{1997}]{Borg}
I.~Borg and P.~Groenen.
\newblock {\em Modern Multidimensional Scaling}.
\newblock New York: Springer-Verlag, 1997.

\bibitem[\protect\citeauthoryear{Chami \bgroup \em et al.\egroup
  }{2020}]{Re_Knowledge}
I.~Chami, A.~Wolf, D.-C. Juan, F.~Sala, S.~Ravi, and C.~Ré.
\newblock Low-dimensional hyperbolic knowledge graph embeddings.
\newblock {\em arXiv preprint arXiv: 2005.00545}, 01 2020.

\bibitem[\protect\citeauthoryear{Cox and Cox}{2001}]{cox}
T.~F. Cox and M.~A.~A. Cox.
\newblock {\em Multidimensional Scaling}.
\newblock London: Chapman Hall, 2001.

\bibitem[\protect\citeauthoryear{Davison}{1983}]{davison}
M.~L. Davison.
\newblock {\em Multidimensional Scaling}.
\newblock New York: Wiley, 1983.

\bibitem[\protect\citeauthoryear{De~Sa \bgroup \em et al.\egroup
  }{2018}]{tradeoff}
C.~De~Sa, A.~Gu, C.~Ré, and F.~Sala.
\newblock Representation tradeoffs for hyperbolic embeddings.
\newblock {\em Proceedings of Machine Learning Research}, 80, 04 2018.

\bibitem[\protect\citeauthoryear{Doreian}{2006}]{csphd}
P.~Doreian.
\newblock Exploratory social network analysis with {P}ajek, {W}. de nooy, {A}.
  {M}rvar, {V}. {B}atagelj. {C}ambridge {U}niversity {P}ress, {N}ew {Y}ork
  (2005).
\newblock {\em Social Networks}, 28:269–274, 07 2006.

\bibitem[\protect\citeauthoryear{Elowitz \bgroup \em et al.\egroup
  }{2002}]{noisegene1}
M.~Elowitz, A.~Levine, E.~Siggia, and P.~Swain.
\newblock Stochastic gene expression in a single cell.
\newblock {\em Science (New York, N.Y.)}, 297:1183--6, 09 2002.

\bibitem[\protect\citeauthoryear{Fonseca \bgroup \em et al.\egroup
  }{2012}]{hyperjeff}
T.~Fonseca, H.~Migon, and M.~Ferreira.
\newblock Bayesian analysis based on the jeffreys prior for the hyperbolic
  distribution.
\newblock {\em Brazilian Journal of Probability and Statistics}, 26, 11 2012.

\bibitem[\protect\citeauthoryear{Gneiting \bgroup \em et al.\egroup
  }{2007}]{calib}
T.~Gneiting, F.~Balabdaoui, and A.~Raftery.
\newblock Probabilistic forecasts, calibration and sharpness.
\newblock {\em Journal of the Royal Statistical Society: Series B (Statistical
  Methodology)}, 69:243 -- 268, 04 2007.

\bibitem[\protect\citeauthoryear{Groenen}{1993}]{Groenen_93}
P.~J.~F. Groenen.
\newblock {\em The Majorization Approach to Multidimensional Scaling: Some
  Problems and Extensions}.
\newblock Liden, The Netherlands: DSWO, 1993.

\bibitem[\protect\citeauthoryear{Hofbauer \bgroup \em et al.\egroup
  }{2016}]{moss}
W.~Hofbauer, L.~Forrest, P.~Hollingsworth, and M.~Hart.
\newblock Preliminary insights from {DNA} barcoding into the diversity of
  mosses colonising modern building surfaces.
\newblock {\em Bryophyte Diversity and Evolution}, 38:1, 04 2016.

\bibitem[\protect\citeauthoryear{Hoff \bgroup \em et al.\egroup }{2002}]{Hoff}
P.~D. Hoff, A.~E. Raftery, and M.~S. Handcock.
\newblock Latent space approaches to social network analysis.
\newblock {\em Journal of the American Statistical Association}, 2002.

\bibitem[\protect\citeauthoryear{Holbrook \bgroup \em et al.\egroup
  }{2021}]{Holbrook}
A.~J. Holbrook, P.~Lemey, G.~Baele, S.~Dellicour, D.~Brockmann, A.~Rambaut, and
  M.~A. Suchard.
\newblock Massive parallelization boosts big {B}ayesian multidimensional
  scaling.
\newblock {\em Journal of Computational and Graphical Statistics}, 30:11--24,
  2021.

\bibitem[\protect\citeauthoryear{Keller-Ressel and Nargang}{2020}]{hydra}
M.~Keller-Ressel and S.~Nargang.
\newblock Hydra: a method for strain-minimizing hyperbolic embedding of
  network- and distance-based data.
\newblock {\em Journal of Complex Networks}, 8, 02 2020.

\bibitem[\protect\citeauthoryear{Klimovskaia \bgroup \em et al.\egroup
  }{2020}]{pseudotime}
A.~Klimovskaia, D.~Lopez-Paz, L.~Bottou, and M.~Nickel.
\newblock Poincaré maps for analyzing complex hierarchies in single-cell data.
\newblock {\em Nature Communications}, 11, 06 2020.

\bibitem[\protect\citeauthoryear{Kosch{\"u}tzki \bgroup \em et al.\egroup
  }{2005}]{centralindice}
D.~Kosch{\"u}tzki, K.~A. Lehmann, L.~Peeters, S.~Richter, D.~Tenfelde-Podehl,
  and O.~Zlotowski.
\newblock {\em Centrality Indices}, pages 16--61.
\newblock Springer Berlin Heidelberg, Berlin, Heidelberg, 2005.

\bibitem[\protect\citeauthoryear{Kruskal}{1964}]{Kruskal}
J.~B. Kruskal.
\newblock Multidimensional scaling by optimizing goodness of fit to a nonmetric
  hypothesis.
\newblock {\em Psychometrika}, 29(1):1--27, 1964.

\bibitem[\protect\citeauthoryear{Liu \bgroup \em et al.\egroup
  }{2022}]{liu_lubold_raftery_mccormick_2022}
B.~Liu, S.~Lubold, A.~Raftery, and T.~McCormick.
\newblock Bayesian hyperbolic multidimensional scaling.
\newblock {\em Presentation at the Joint Statistical Meetings, Washington,
  D.C.}, August 2022.

\bibitem[\protect\citeauthoryear{Lubold \bgroup \em et al.\egroup
  }{2020}]{Lubold2020}
S.~Lubold, A.~Chandrasekhar, and T.~McCormick.
\newblock Identifying the latent space geometry of network models through
  analysis of curvature.
\newblock {\em arXiv preprint arXiv:2012.10559}, 2020.

\bibitem[\protect\citeauthoryear{Lukk \bgroup \em et al.\egroup }{2010}]{lukk}
M.~Lukk, M.~Kapushesky, J.~Nikkilä, H.~Parkinson, A.~Goncalves, W.~Huber,
  E.~Ukkonen, and A.~Brazma.
\newblock A global map of human gene expression.
\newblock {\em Nature Biotechnology}, 28:322--4, 04 2010.

\bibitem[\protect\citeauthoryear{MacKay}{1989}]{Mackay}
D.~MacKay.
\newblock Probabilistic multidimensional scaling: An anisotropic model for
  distance judgements.
\newblock {\em Marketing Science}, 5:325--334, 1989.

\bibitem[\protect\citeauthoryear{Nagano \bgroup \em et al.\egroup
  }{2019}]{wrapnorm}
Y.~Nagano, S.~Yamaguchi, Y.~Fujita, and M.~Koyama.
\newblock A wrapped normal distribution on hyperbolic space for gradient-based
  learning.
\newblock {\em arXiv preprint arXiv: 1902.02992}, 2019.

\bibitem[\protect\citeauthoryear{Nickel and Kiela}{2017}]{poinword}
M.~Nickel and D.~Kiela.
\newblock Poincaré embeddings for learning hierarchical representations.
\newblock {\em arXiv preprint arXiv: 1705.08039}, 05 2017.

\bibitem[\protect\citeauthoryear{Oh and Raftery}{2001}]{oh2001bayesian}
M.-S. Oh and A.~E. Raftery.
\newblock Bayesian multidimensional scaling and choice of dimension.
\newblock {\em Journal of the American Statistical Association},
  96(455):1031--1044, 2001.

\bibitem[\protect\citeauthoryear{Okamoto \bgroup \em et al.\egroup
  }{2008}]{centrality}
K.~Okamoto, W.~Chen, and X.-Y. Li.
\newblock Ranking of closeness centrality for large-scale social networks.
\newblock volume 5059, pages 186--195, 06 2008.

\bibitem[\protect\citeauthoryear{Oleksiak \bgroup \em et al.\egroup
  }{2002}]{noisegene2}
M.~Oleksiak, G.~Churchill, and D.~Crawford.
\newblock Variation in gene expression within and among natural populations.
\newblock {\em Nature Genetics}, 32:261--6, 11 2002.

\bibitem[\protect\citeauthoryear{Plummer \bgroup \em et al.\egroup
  }{2006}]{coda}
M.~Plummer, N.~Best, K.~Cowles, and K.~Vines.
\newblock {CODA}: Convergence diagnosis and output analysis for {MCMC}.
\newblock {\em R News}, 6(1):7--11, 2006.

\bibitem[\protect\citeauthoryear{Praturu and
  Sharpee}{2022}]{Praturu2022.10.12.511940}
A.~Praturu and T.~Sharpee.
\newblock {A} {B}ayesian approach to hyperbolic multi-dimensional scaling.
\newblock {\em bioRxiv}, 2022.

\bibitem[\protect\citeauthoryear{Raftery and
  Lewis}{1992}]{raftery1992practical}
A.~E. Raftery and S.~M. Lewis.
\newblock Comment: one long run with diagnostics: implementation strategies for
  {M}arkov chain {M}onte {C}arlo.
\newblock {\em Statistical science}, 7(4):493--497, 1992.

\bibitem[\protect\citeauthoryear{Raftery \bgroup \em et al.\egroup
  }{2012}]{lscasecontrol}
A.~Raftery, X.~Niu, P.~Hoff, and K.~Y. Yeung.
\newblock Fast inference for the latent space network model using a
  case-control approximate likelihood.
\newblock {\em Journal of Computational and Graphical Statistics}, 21, 10 2012.

\bibitem[\protect\citeauthoryear{Raj and Oudenaarden}{2008}]{noisegene3}
A.~Raj and A.~Oudenaarden.
\newblock Nature, nurture, or chance: Stochastic gene expression and its
  consequences.
\newblock {\em Cell}, 135:216--26, 11 2008.

\bibitem[\protect\citeauthoryear{Smith \bgroup \em et al.\egroup
  }{2019}]{smith2019geometry}
A.~L. Smith, D.~M. Asta, C.~A. Calder, et~al.
\newblock The geometry of continuous latent space models for network data.
\newblock {\em Statistical Science}, 34(3):428--453, 2019.

\bibitem[\protect\citeauthoryear{Stai \bgroup \em et al.\egroup
  }{2016}]{hyperclose1}
E.~Stai, V.~Karyotis, and S.~Papavassiliou.
\newblock A hyperbolic space analytics framework for big network data and their
  applications.
\newblock {\em IEEE Network}, 30(1):11--17, 2016.

\bibitem[\protect\citeauthoryear{Takane and Caroll}{1981}]{Takane}
Y.~Takane and J.~D. Caroll.
\newblock Nonmetric maximum likelihood multidimensional scaling from
  directional rankings of similarities.
\newblock {\em Psychometrika}, 46:389--405, 1981.

\bibitem[\protect\citeauthoryear{Turker and Balcisoy}{2013}]{hyperclose2}
U.~Turker and S.~Balcisoy.
\newblock A visualization technique for large temporal social network datasets
  in hyperbolic space.
\newblock {\em Journal of Visual Languages \& Computing}, 25, 01 2013.

\bibitem[\protect\citeauthoryear{Zachary}{1976}]{karate}
W.~Zachary.
\newblock An information flow model for conflict and fission in small groups.
\newblock {\em Journal of anthropological research}, 33, 11 1976.

\bibitem[\protect\citeauthoryear{Zhou and Sharpee}{2021}]{sharpee}
Y.~Zhou and T.~Sharpee.
\newblock Hyperbolic geometry of gene expression.
\newblock {\em iScience}, 24:102225, 02 2021.

\end{thebibliography}

\newpage
\begin{appendices}
\setcounter{figure}{0}    
\renewcommand\thefigure{\thesection.\arabic{figure}}

\section{Reasoning of the Prior Choice}
\label{sec: why_Nagano}
Note that, since each $\boldsymbol x_i$ is defined on $\mathbb{H}^p(\kappa)$, we cannot apply a usual multivariate normal prior on $\boldsymbol X$ as in \cite{oh2001bayesian}, which is originally defined on the Euclidean geometry. It is tempting to consider distributions defined on the hyperbolic space as substitutes of the multivariate normal distribution, such as the distributions described in \cite{hyperjeff} and \cite{wrapnorm}, and directly specify the prior on $\boldsymbol X$. However, since the hyperbolic space is centered around $\boldsymbol \mu_0^p = (1, 0, \cdots, 0) \in \mathbb{H}^p \subset \mathbb{R}^{p + 1}$ instead of the Euclidean origin $(0, \cdots, 0) \in  \mathbb{R}^{p + 1}$, directly impose a hyperbolic prior on $\boldsymbol X$ will generally lead to an asymmetric proposal function, and restrict us from applying the random walk Metropolis-Hastings sampling algorithm described in \cite{oh2001bayesian}, which uses a symmetric proposal function, and is more tractable and computationally efficient.

\section{Curvature Estimation through Stress Minimization}
\label{sec: kappa}

In this section, we define an estimate of the curvature of the $\mathbb{H}^p(\kappa)$ given (potentially) noisy dissimilarities between points on $\mathbb{H}^p(\kappa)$. An advantage of the estimator we now propose is that it does not depend on knowing the dimension $p$ of the space. For any curvature value $\kappa$, we let $\boldsymbol x_i(\kappa)$ for $i = 1, \dotsc,n$ denote the set of embedding coordinates obtained from BHMDS or \texttt{hydraPlus} computed using the curvature value $\kappa$, and we let $d_{ij}(\kappa)$ be the distance between $\boldsymbol x_i(\kappa)$ and $\boldsymbol x_j(\kappa)$.  Our estimate $\hat \kappa$ is then the estimate that minimizes the stress between the observed dissimilarities $\{\widehat \delta_{ij}\}$ and the distances $\{d_{ij}(\kappa)\}$. That is, we set 
\begin{equation*}
    \hat \kappa = \arg \min_{\kappa}     \text{stress}\left(\{d_{ij}(\kappa)\}, \{\widehat{\delta}_{ij}\}\right) =: \arg \min_{\kappa} \sqrt{\frac{\sum_{i<j}\left(d_{i j}(\kappa)-\widehat{\delta}_{i j}\right)^{2}}{\sum_{i<j} d_{i j}(\kappa)^{2}}} \;.
\end{equation*}
\cbstart
In general, we find BHMDS yields better curvature estimates than \texttt{hydraPlus}, thus we recommend using BHMDS for curvature estimation if computation power permits.  
\cbend
This is not the only way to estimate the curvature of $\mathbb{H}^p(\kappa)$ given (potentially) noisy dissimilarity data. \cite{Lubold2020}, for example, proposed a different estimate of curvature and proved it is consistent as the error in $\{\widehat \delta_{ij}\}$ goes to zero. But we found in simulations that the above estimator outperforms the estimate in \cite{Lubold2020} for the noise levels we consider in this work. 

\cbstart
\section{Identifiability and Hyperbolic Visualization}

In general, hyperbolic visualization refers to representing objects around the center of an (often) two-dimensional Poincare disk. Yet, due to the nature of hyperbolic space, the volume of hyperbolic space is not constant from the origin, and it is possible to transform the hyperbolic coordinates through Procrustean operations and represent the objects using only part of the Poincare disk, with the distances between coordinates to be invariant. One may refer to \cite{pseudotime} for more detail. This suggests there is no unique way in visualizing hyperbolic data.

In BHMDS, we consider a hyperbolic wrapped normal distribution centered at the hyperbolic origin as the prior of the coordinates. Such a prior choice indicates we attempt to embed objects at the center of the hyperbolic space. Specifically, BHMDS will try to embed objects with equal and small distances to all others around the hyperbolic origin and to embed objects with large distances to others away.

Technically, our Bayesian estimate of $\boldsymbol{X}$ is not precisely centered at the origin as specified in the prior. A set of $\boldsymbol{X}$ is said to be centered at the origin if its inverse transformation $\boldsymbol V = T^{-1}(\boldsymbol X)$ on the Euclidean space has mean zero and diagonal covariance matrix. In Euclidean MDS, it is possible to center the Euclidean coordinates by Procrustean operation (see postprocessing in \cite{oh2001bayesian}). In the hyperbolic space, we cannot use Procrustean operations on $\boldsymbol{V}$ to center $\boldsymbol{X}$, as the transformation $T$ is not invariant to Procrustean operations so that the distances are not preserved after centering $\boldsymbol{V}$. Still, from simulations and the network example in the newly added Section \ref{sec: network}, BHMDS consistently embed objects by their distances to the origin and identifies nodes with high centrality. This suggests BHMDS embeds as theoretically expected.

In other applications, practitioners identify a certain object as the root of structure. For instance, in genomic study, embryonic stem cells can be viewed as the root of the differentiation process. The choices of root are often subjective to expertise knowledge, so that visualization is highly customized and changes with respect to different choices of the root. \cite{pseudotime} describes a transformation that centers the hyperbolic coordinates by setting the root object at the hyperbolic origin while preserving the hyperbolic distances. Practitioners can use such transformation to recenter the BHMDS estimate accordingly for visualization.

Thus, if the goal is to identify nodes that is central and close to all others, we may directly visualize the BHMDS result on the Poincare disk, as in Section \ref{sec: network}. Otherwise, one will need to transform the estimates before visualization as in \cite{pseudotime}. 

Additionally, when the coordinates are centered with respect to the root, their distances to the origin will change accordingly. This suggests that, if the distances to the origins are of practical interest, we need to postprocess the MCMC samples by the transformation before calculating the distances to the origin. In the example in \ref{sec: gene}, we have not identified a cell type that is commonly recognized as the root of differentiation, but we still observe the more de-differentiated breast cancer cell and the less differentiated germ cell neoplasm embedded around the hyperbolic origin. This indicates postprocessing is not necessary in the example.
\cbend
\section{MCMC Convergence}

In this section, we provide typical trace plots from our simulations and data sets. We plot the MCMC proposed $\delta_{ij}$ and $\sigma$ values over iterations $t$ for the MCMC simulation in Section \ref{sec: simulations}. Specifically, to comprehensively investigate the convergence of $\delta_{ij}$, we ran a BHMDS simulation for sample size $n = 200$, hyperbolic dimension $p = 2$, error size $\sigma = 1$, with 20,000 MCMC iterations and burn-in of 3,000 iterations, randomly pick ten $\delta_{ij}$ entries from the dissimilarity matrix, and record their MCMC sampled values after the MCMC burn-in.  We provide the trace plots below.  We also assessed convergence using the diagnostic of~\citet{raftery1992practical}. We analyze the traces using the default \texttt{raftery.diag()} function from the \texttt{coda} R package \citep{coda}, and summarize the results in Table \ref{tab: diag}. Our choice of the total number of MCMC iterations is close to the average of the suggested total number of MCMC iterations, $\bar{N} = 21596$ iterations $Nmin = 3746$. This suggests that around 20,000 MCMC iterations are enough to estimate the parameters of interest in BHMDS.  

\begin{figure}[ht]
    \centering
    \includegraphics[width =\linewidth]{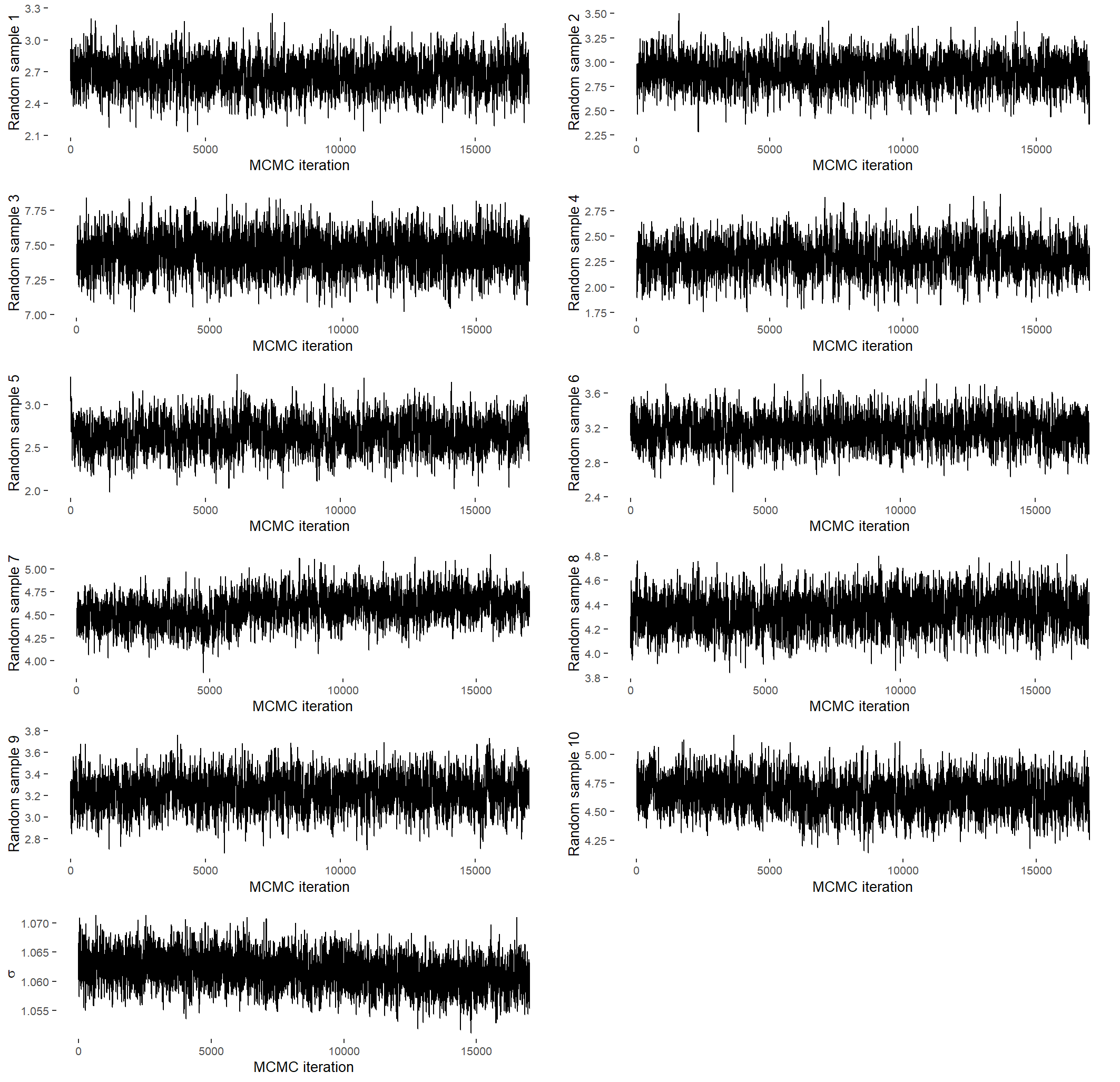}
    \caption{\footnotesize{Trace plots of ten randomly sampled MCMC proposed $\delta_{ij}$ and $\sigma$ values over iteration $t$ for MCMC simulation in Section \ref{sec: simulations} with $n = 200$, $p = 2$, $\sigma = 1$.}}
    \label{fig: trace1}
\end{figure}

\end{appendices}

\begin{table} 
\caption{\footnotesize{Summary of the \texttt{raftery.diag()} MCMC diagnosis for traces of the ten $\delta_{ij}$ random samples and error size $\sigma$. $M$ is small for all traces, indicating we burn-in enough of the MCMC samples. Our choice of the total number of MCMC iterations is close to the suggested total number of MCMC iterations $N$'s, and greatly exceeds the suggested minimum number of iterations $Nmin$, indicating the proposed MCMC sampler mixes well with $\sim 20000$ MCMC iterations.}}
\vspace{20pt}
\centering
\renewcommand{\arraystretch}{0.8}
 \begin{tabular}{c c c c c} 
 \hline
&Burn-in (M) & Total (N) & Lower bound (Nmin) &  Dependence factor (I)\\
 \hline
Sample 1 &
21 &
24222 &
3746&
6.47 \\
Sample 2 &
18 &
19083 &
3746 &
5.09 \\
Sample 3 & 
18 &
20487 &
3746&
5.47 \\
Sample 4 & 
24 & 
24474 & 
3746 & 
6.53 \\
Sample 5 & 
20 & 
22194 & 
3746 & 
5.92 \\
Sample 6 & 
18 & 
20061 & 
3746 & 
5.36 \\
Sample 7 & 
18 & 
21762 & 
3746 & 
5.81 \\
Sample 8 & 
20 & 
22152 & 
3746 & 
5.91 \\
Sample 9 & 
20 & 
23348 & 
3746 & 
6.23 \\
Sample 10 & 
18 & 
21612 & 
3746 & 
5.77 \\
$\sigma$& 
18 & 
18158 & 
3746 & 
4.85 \\

 \hline
 \end{tabular}
 \vspace{10pt}
 \label{tab: diag}
\end{table}

\end{document}